\documentclass[ journal]{IEEEtran}
\usepackage{verbatim}
\usepackage{tikz}
\usetikzlibrary{spy}
\usetikzlibrary{arrows,shapes}
\usepackage{cite}
\usepackage{amsmath,amssymb,latexsym,bm}
\usepackage{paralist}
\usepackage{graphicx}
\usepackage{float}
\usepackage{algorithmic}
\usepackage{algorithm}
\usepackage{cases}
\usepackage{subfigure}
\usepackage{caption2}
\usepackage{epstopdf}
\usepackage{makecell,multirow,diagbox}


\usepackage{xcolor}
\usepackage{tikz}
\usetikzlibrary{spy}
\usepackage{wrapfig}
\usetikzlibrary{arrows}
\tikzstyle{block}=[draw opacity=0.7,line width=1.4cm]
\tikzstyle{process} = [rectangle, minimum width=3cm, minimum height=1cm, text centered, draw=black, fill = yellow!50]

\usepackage[colorlinks=true]{hyperref}
\hypersetup{urlcolor=blue, citecolor=red,linkcolor=blue,}

\begin{document}
\title{Low-Dose CT with Deep Learning Regularization via Proximal Forward Backward Splitting
}
\author{Qiaoqiao Ding\IEEEauthorrefmark{1}, Gaoyu Chen, Xiaoqun Zhang, Qiu Huang,  Hui Ji\IEEEauthorrefmark{1} and  Hao Gao\IEEEauthorrefmark{1}%
	\IEEEcompsocitemizethanks{
		\IEEEcompsocthanksitem Q. Ding\IEEEauthorrefmark{1} (e-mail: matding@nus.edu.sg) and H. Ji\IEEEauthorrefmark{1} (e-mail: matjh@nus.edu.sg) are with Department  of Mathematics, National University of Singapore, 
		 119076, SINGAPORE
		\IEEEcompsocthanksitem G. Chen  and Q. Huang are with School of Biomedical Engineering, Shanghai Jiaotong University, Shanghai, 200240, CHINA 
		\IEEEcompsocthanksitem X. Zhang  is  with Institute of Natural Sciences and School of Mathematical Sciences, Shanghai Jiaotong University, Shanghai, 200240, CHINA 
		\IEEEcompsocthanksitem  H. Gao\IEEEauthorrefmark{1} (e-mail: hao.gao.2012@gmail.com)  is with  Department of Radiation Oncology, Winship Cancer Institute of Emory University, Atlanta, Georgia,  30322, USA
	}
}
\maketitle
\thispagestyle{empty}
\begin{abstract}
	 Low dose X-ray computed tomography (LDCT) is desirable for reduced patient dose. This work develops image reconstruction methods with  deep learning (DL) regularization for LDCT. Our methods are based on unrolling of  proximal forward-backward splitting (PFBS) framework with data-driven image regularization via deep neural networks.
	 In contrast with PFBS-IR that utilizes standard data fidelity updates via iterative reconstruction (IR) method, PFBS-AIR involves preconditioned data fidelity updates that
	 fuse  analytical reconstruction (AR)  method and IR in a synergistic way,  \emph{i.e.}, fused analytical and iterative reconstruction (AIR).
	 The results suggest that DL-regularized methods (PFBS-IR and PFBS-AIR) provided better reconstruction quality from conventional wisdoms (AR or IR), and DL-based postprocessing method (FBPConvNet). In addition, owing to AIR, PFBS-AIR  noticeably outperformed PFBS-IR.
\end{abstract}
\section{Introduction}
\IEEEPARstart{L}{ow} dose X-ray computed tomography (LDCT) is desirable for reduced patient dose. However, standard analytical reconstruction (AR) methods  often yield low-dose image artifacts due to reduced signal-to-noise ratio.
Thus  iterative reconstruction (IR) methods  have been actively explored  to  reduce low-dose artifacts for LDCT.
 In the IR method, the reconstruction problem is often formulated as an optimization problem with a data fidelity term and a regularization term. A popular category of IR in the last decades is via the sparsity regularization, such as total variation \cite{zhang2005total,sidky2008image,chen2008prior}, wavelet tight frames \cite{jia2011gpu,gao20124d}, nonlocal sparsity \cite{jia20104d},
 and low-rank models \cite{gao2011robust,gao2011multi,cai2014cine,chen2015synchronized,gao2018principal}.

Recently, deep learning (DL) methods have been studied extensively for CT image reconstruction. A major difference  among  different  various DL-based postprocessing methods lies in the choice of  network architecture,  \emph{e.g.},  residual network \cite{chen2017low,han2016deep,li2017low}, U-net \cite{han2016deep,jin2017deep} and  generative adversarial network (GAN)/Wasserstein-GAN \cite{wolterink2017generative,yang2018low}. Instead of  deep neural network (DNN) directly based on reconstructed images in the image domain, DNN based  on the transform coefficients of  reconstructed images in the transform domain can be designed for further improvement, \emph{e.g.}, in the  wavelet transform \cite{kang2017deep,gu2017multi}.

Alternative to DL-based image postprocessing, DL can be integrated with image reconstruction. This is often done by unrolling some iterative optimization schemes and replacing conventional  regularization methods with convolutional neural networks (CNN).  In  \cite{sun2016deep}, Yang et al. proposed a DL-regularized method via alternating direction method of multipliers, ADMM-net, for magnetic resonance (MR) image reconstruction.
In \cite{mardani2017deep,mardani2017recurrent}, Mardani et al. proposed the proximal methods for MR imaging using GAN.
Similar unrolling methods were also recently proposed for CT  reconstruction.
In \cite{chen2018learn}, Chen  et al. developed the gradient descent based IR method that used CNN to learn image regularization  for sparse view CT reconstruction.
In \cite{adler2018learned}, Adler et al. proposed  an image reconstruction method  by
unrolling a proximal primal-dual optimization method, where the proximal operators were replaced with CNN.
In \cite{gupta2018cnn}, Gupta et al. replaced the projector in a projected gradient descent  proposed method with a CNN. 
In \cite{he2018optimizing}, He et al. proposed a DL-based IR method by unrolling the framework of ADMM for LDCT.

The work will explore DL-regularized image reconstruction method for LDCT using the framework of proximal forward-backward splitting (PFBS). Our method will unroll the optimization by PFBS with data-driven image regularization learned by DNN. To further improve image reconstruction quality, a preconditioned PFBS version will be used with fused analytical and iterative reconstruction (AIR) \cite{gao2016fused}. Thus, the proposed method will integrate AR, IR and DL using the PFBS framework for LDCT. 

\section{Method}
\label{Method}
\subsection{Preliminaries}
\label{Preliminaries}
CT image reconstruction problem can be formulated  as solving an ill-posed linear system:
\begin{equation}
\bm{y}=\bm{A}\bm{x}+\bm{n}.
\end{equation}
Here $\bm{x}$ denotes the attenuation map with  $x_j$ being the linear attenuation coefficient in the $j$-th pixel for $j =1,\cdots,N_p$ and  $N_p$ denotes the total number of pixels; 
$\bm{y}$ represents the measured projection after correction and log transform. The matrix 
$\bm{A}$ is the $N_d \times N_p$ system matrix with entries $a_{ij}$, and $[\bm{A}\bm{x}]_i=\sum_{j=1}^{N_p}a_{ij}x_{j}$ denotes the line integral of the attenuation map $\bm{x}$ along the $i$-th X-ray with $i=1\cdots N_d$. CT image reconstruction problem  is to recover the unknown image $\bm{x}$, provided the system matrix $\bm{A}$  and the projection data $\bm{y}$ in the presence of 
measurement noise $\bm{n}$.

Although AR is fast, it suffers from the low-dose artifacts. Alternatively, IR for   image  reconstruction with flexible models for both data fidelity and image regularization. In its general form, IR is  formulated as solving the optimization problem:
\begin{equation}
\mathcal{R}_\lambda(\bm{y})=\arg\min_{\bm{x}}\mathcal{L}(\bm{A}\bm{x},\bm{y})+\lambda \psi(\bm{x}).
\label{InvPro}
\end{equation}
Here the first term $\mathcal{L}(\bm{A}\bm{x},\bm{y})$  is the data  fidelity term, where $\bm{x}$ and $\bm{y}$ are elements in appropriate function space $X$ and $Y$ and the  forward operator is a mapping $\bm{A}:X\rightarrow Y$;
the second term
$\psi(\bm{x})$ is  the image regularization  that imposes certain  prior knowledge on the image $\bm{x}$.
$\mathcal{R}_\lambda:Y \rightarrow X $ denotes the  reconstruction operator with the regularization  parameter $\lambda$.
For  Gaussian modeled noise $\bm{n}$ weighted by $\bm{W}$, the data fidelity term is given by 
\begin{equation}
\mathcal{L}(\bm{A}\bm{x},\bm{y})=\frac{1}{2}\|\bm{A}\bm{x}-\bm{y}\|_{\bm{W}}^2. 
\end{equation}

\subsection{Proximal Forward Backward  Splitting}
Operator splitting methods have been extensively studied in the optimization community, \emph{e.g.} \cite{combettes2005signal,eckstein1992douglas,glowinski1989augmented}. They aim to minimize the sum of two convex functions
\begin{equation}
\min_{\bm{x}} \mathcal{L}(\bm{x})+\lambda \psi(\bm{x}).
\label{prob}
\end{equation}
The forward-backward technique based on the proximal operator for general signal recovery tasks was  introduced by Combettes and Wajs. 
The proximal operator of a convex functional $\psi$, which was originally introduced by Moreau in \cite{moreau1962fonctions}, is defined as
\begin{equation}
\mathbf{Prox}_{\lambda\psi}(\cdot)=\arg\min_{\bm{x}}\lambda\psi(\bm{x}) +\frac{1}{2}\|\bm{x}-\cdot\|_2^2.
\label{PFBS}
\end{equation}
By classic arguments of convex analysis, the solution  of (\ref{prob}) satisfies the condition
\begin{equation}
0\in \partial \mathcal{L}(\bm{x})+ \lambda\partial \psi(\bm{x}).
\end{equation}
For a positive number $\alpha$, we obtain:
\begin{equation}
0\in(\bm{x}+\alpha\lambda\partial \psi(\bm{x}))- (\bm{x}-\alpha\partial \mathcal{L}(\bm{x})).
\end{equation}
This lead to a forward and backward splitting algorithm:
\begin{equation}
\bm{x}^{k+1}=\mathbf{Prox}_{\alpha\lambda\psi}(\bm{x}^{k}-\alpha\partial \mathcal{L}(\bm{x}^{k})).
\label{FBS}
\end{equation}

The forward and backward splitting algorithm (\ref{FBS}) is	equivalent to
	\begin{subequations}
		\begin{numcases}
		{} \bm{x}^{k+\frac{1}{2}}=\arg\min_x \mathcal{L}(\bm{x}),\\
		\bm{x}^{k+1}=\arg\min_x \alpha\lambda\psi(\bm{x})+ \frac{1}{2}\| \bm{x} -\bm{x}^{k+\frac{1}{2}}\|,
		\end{numcases}
	\end{subequations}
	where the first subproblem is solved by  gradient descent method with initial value  $ \bm{x}^k$ and step size $\alpha$,
	$$\bm{x}^{k+\frac{1}{2}}=\bm{x}^k-\alpha \bm{A}^T(\bm{A }\bm{x}^k-\bm{y}). $$ 
	Inspired by Newton's method, we consider the preconditioned gradient descent \cite{zhang2010bregmanized} in reconstruction problem (\ref{InvPro}):
	$$\bm{x}^{k+\frac{1}{2}}=\bm{x}^k-\alpha \bm{A}^+(\bm{A }\bm{x}^k-\bm{y}), $$
	where $\bm{A }^+$  is the pseudo-inverse of $\bm{A }$. The operator $\bm{A}^+\bm{A}$ is an orthogonal projector onto the range space of 
	$\bm{A }^+$

Thus,   (\ref{InvPro}) can be solved by the following two-step algorithm
\begin{subequations}
	\label{MPFBS}
	\begin{numcases}
	{}
	\bm{x}^{k+\frac{1}{2}}=\bm{x}^{k}-\alpha \bm{A}^+(\bm{A }\bm{x}^{k}-\bm{y}), \label{gdstep} \\
	\bm{x}^{k+1}=\mathbf{Prox}_{\alpha\lambda\psi}(\bm{x}^{k+\frac{1}{2}} ),\label{proxstep}
	\end{numcases}
\end{subequations}
where $\bm{A}^+$ is the approximate of the pseudo inverse of $\bm{A}$.
For example, if $\bm{AA}^T$ and/or  $\bm{A}^T\bm{A}$  do not exist,  $\bm{A}^+$ is set 
as the Moore-Penrose pseudo inverse of $\bm{A}$,  \emph{i.e.}
$$\lim_{\varepsilon\rightarrow0}(\bm{A}^T\bm{A}+\varepsilon\bm{I})^{-1}\bm{A}^T=\lim_{\varepsilon\rightarrow0}\bm{A}^T(\bm{A}^T\bm{A}+\varepsilon\bm{I})^{-1}=\bm{A}^+.$$
More precisely, $\bm{A}^+$ can be set as AR operator.
The convergence property of the preconditioned PFBS algorithm (\ref{MPFBS}) have been given in \cite{gao2016fused}. 
\subsection{DL-Regularized PFBS}
\label{Learn}

We present a new image reconstruction method that unrolls PFBS with data-driven image regularization via DNN for LDCT.

The aim of unrolling is to find a DNN architecture 
$\mathcal{R}_\theta: Y\rightarrow X$ that is suitable for  approximating the operator $\mathcal{R}_\lambda : Y\rightarrow X $  defined via PFBS iterative scheme.
Unrolled iterative scheme of the preconditioned PFBS algorithm consists of in two steps. 
Firstly, by unrolling the $k$-th  gradient-descent step for data fidelity  of  \eqref{MPFBS}, we have 
\begin{align}
\bm{x}^{k+\frac{1}{2}}=\Lambda_{\theta^1_k} ( \bm{x}^{k}, \bm{A}^+(\bm{A }\bm{x}^{k}-\bm{y})),
\end{align}
where $\Lambda_{\theta^1_k}:X\times X\rightarrow X$ is a learned  operator. 
The learned updating in a gradient descent scheme, $\Lambda_{\theta^1_k}( x,z)=x- \theta^1_k z$, implies
the step length is also learned and varies with iterations.

Secondly, by  replacing the proximal operator with a learned operator, it yields:
\begin{align}
	\bm{x}^{k+1}:=\Lambda_{\theta^2_k}(\bm{x}^{k+\frac{1}{2}}, \bm{x}_k ) \label{CNN}. 
\end{align}
In equation \eqref{CNN}, $\Lambda_{\theta^2_k}:X\times X^k\rightarrow X$  is a CNN with learned parameters to model $\Lambda_{\theta^2_k}$ that replaces the proximal of  \eqref{MPFBS} and
$\bm{x}_k = (\bm{x}^{\frac{1}{2}},\bm{x}^{1+\frac{1}{2}},\cdots\bm{x}^{k-\frac{1}{2}})\in X^k$.
In reality, the proximal operator in \eqref{MPFBS} is a mapping from $\bm{x}^{k+\frac{1}{2}}$ to $\bm{x}^{k+1}$.
By the formulation of \eqref{proxstep},  we can naturally  set \eqref{CNN} as  
\begin{equation}
{\rm{CNN}}(\cdot,\theta^2_{k}): \bm{x}^{k+\frac{1}{2}} \rightarrow \bm{x}^{k+1}.
\end{equation}
In fact, the   applied CNN is modified by concatenating all previous estimates of the latent image as the input. 
We replace ${\rm{CNN}}(\bm{x}^{k+\frac{1}{2}},\theta^2_{k} )$ with a densely-connected  ${\rm{CNN}}([\bm{x}^{\frac{1}{2}},\bm{x}^{1+\frac{1}{2}},\cdots\bm{x}^{k+\frac{1}{2}}],\theta^2_{k} )$ \cite{huang2017densely,zhang2018sparse}, which was utilized in our previous work for  sparse-data  CT \cite{chen2019airnet} and shown to outperform the standard CNN.
The problem of vanishing gradient  can be addressed by the modification.
Thus, we replace the proximal operator with  learnable parameters as \eqref{CNN}.

To summarize, the preconditioned PFBS based unrolling scheme is as follows.
\begin{subequations}
	\label{subpro}
	\begin{numcases}
	{}\bm{x}^{k+\frac{1}{2}}=\Lambda_{\theta^1_k}( \bm{x}^{k}, \bm{A}^+(\bm{A} \bm{x}^{k}-\bm{y})),\label{gd} \\
	\bm{x}^{k+1}=\Lambda_{\theta^2_k}(\bm{x}^{k+\frac{1}{2}},\bm{x}_k).	\label{cn}
	\end{numcases}
\end{subequations}
In the iteration  scheme \eqref{subpro}, $\bm{x}^{k+\frac{1}{2}}$ and $\bm{x}^{k+1}$ are two intermediate iterates, where $\bm{x}^{k+\frac{1}{2}}$  is from the data fidelity specific to the reconstruction problem, and  $\bm{x}^{k+1}$ is from the learning that is data-driven. In each iteration, during the 
gradient descent step \eqref{gd}, the image is projected to the data domain to form the data residual, and then backprojected to the image domain by an AR operator, $\bm{A}^+$, to form a residual image, which is then weighted together with previous image iterate to generate the current image iterate; during the proximal step \eqref{cn}, all previous estimates of the image are concatenated to learn the image priors from the training data.

Then  the truncated scheme after $\rm{K}$ iterates amounts to defining $\mathcal{R}_\theta :Y\rightarrow X$ with $\theta=(\theta^1_0,\cdots\theta^1_{K-1}, \theta^2_0,\cdots\theta^2_{K-1}) $ as
\begin{align}
\mathcal{R}_\theta(\bm{y}):=\bm{x}^K= (\Lambda_{\theta^2_{K-1}} \circ \Lambda_{\theta^1_{K-1}}\cdots \Lambda_{\theta^2_0} \circ \Lambda_{\theta^1_0} )(\bm{x}^0, \bm{A }\bm{x}^{0}-\bm{y}).
\end{align}
The initial value  $\bm{x}^0$ is set as $\bm{A}^+\bm{y}$.
There are totally $\rm{K}$ stages  and each stage corresponds to a outer iteration in the scheme \eqref{MPFBS}. See
Fig.~\ref{fig:1} for the  diagram of the proposed method, named PFBS-IR or PFBS-AIR, for reconstruction  LDCT images.
\tikzstyle{startstop} =[draw, thin,draw =blue, fill=blue!20]
\tikzstyle{format} = [draw, thin, draw =blue,fill=blue!20]
\tikzstyle{process} = [rectangle, minimum width=3cm, minimum height=1cm, text centered, draw=blue, fill = blue!20]
\tikzstyle{arrow} = [->,>=stealth,blue,line width=0.05cm]
\tikzstyle{Out} = [draw, thin,draw =blue, fill=blue!20]
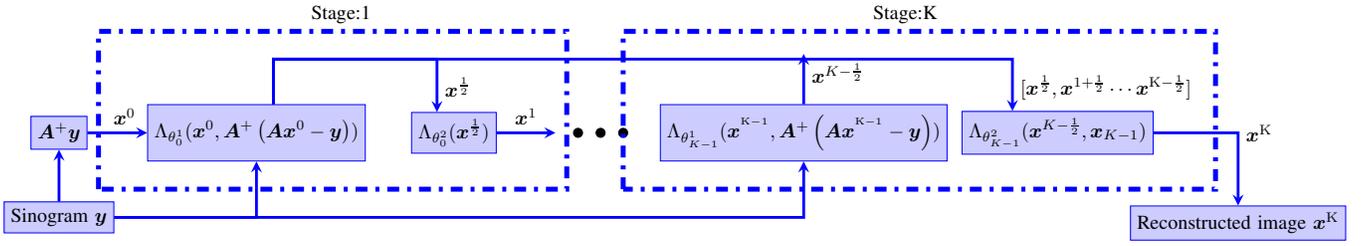
\begin{figure*}[!htp]
	\scalebox{0.75}{
		\centering
		\begin{tikzpicture}[node distance=2cm]
		\path[->] node[format] (Inv0) {$\bm{A}^+\bm{y}$};
		\path[->] node[startstop, below of=Inv0,yshift=-0.5cm] (Input) at (0,1) {Sinogram $\bm{y}$};
		\path[->] node[process, right of=Inv0,xshift=1.5cm] (INV1)  {$\Lambda_{\theta^1_0}(\bm{x}^{0}, \bm{A}^+\left(\bm{Ax}^{0}-\bm{y}\right))$};
		\path[->] node[format, right of=INV1,xshift=1.5cm] (CNN1)  {$\Lambda_{\theta^2_0}(\bm{x}^{\frac{1}{2}} )$};
		\path[->] node[process, right of=CNN1,xshift=4.2cm] (INV2) {$\Lambda_{\theta^1_{K-1}}(\bm{x}^{^{{\rm{K}}-1}},\bm{A}^+\left(\bm{Ax}^{^{{\rm{K}}-1}}-\bm{y}\right))$};
		\path[->] node[format, right of=INV2,xshift=2.5cm] (CNN2) {$\Lambda_{\theta^2_{K-1}}(\bm{x}^{K-\frac{1}{2}},\bm{x}_{K-1})$ };
		\path[->] node[Out, below of=CNN2,yshift=0.4cm,xshift=3.2cm] (OutPut) {Reconstructed image $\bm{x}^{{\rm{K}}}$ };
		\draw [arrow] (Input) -- (Inv0);
		\draw [arrow] (Inv0) -- node [above,black] {~~$\bm{x}^{0}$} (INV1);
		\draw[arrow] ([yshift=0.0cm, xshift=0.3cm]INV1.north)--++(0, 0.8cm)-|node [right,black,xshift=-0.0cm,yshift=-0.5cm] {$\bm{x}^{\frac{1}{2}}$}([yshift=0.0cm, xshift=-0.3cm]CNN1.north);
		\draw[arrow] ([yshift=0.0cm, xshift=0.0cm]CNN1.east)--node [above,black] {$\bm{x}^{1}$ }(8.8, 0.0cm);
		\draw [fill] (9.2,0) circle [radius=.08];
		\draw [fill] (9.6,0) circle [radius=.08];
		\draw [fill] (10.0,0) circle [radius=.08];
		\draw [arrow] (CNN2) -|node [right,black] {$\bm{x}^{{\rm{K}}}$ } (OutPut);
		\draw [arrow] (Input) -|node [right] { } (INV1);
		\draw [arrow] (Input) -|node [right] { } (INV2);
		\draw[line width=2.5pt,blue,dash pattern=on 6pt off 4pt on 2pt off 4pt,-] (0.7,-1)--(0.7,1.8)--(9.0,1.8)--(9.0,-1)--(0.7,-1);
		\draw[line width=2.5pt,blue,dash pattern=on 6pt off 4pt on 2pt off 4pt,-] (10.0,-1)--(10.0,1.8)--(20.5,1.8)--(20.5,-1)--(10.0,-1);
		\node at	(5,2.1)	{Stage:1};
		\node at	(15,2.1){Stage:K};
		\draw[arrow] ([yshift=0.0cm, xshift=0.3cm]INV1.north)--++(0, 0.8cm)-|node [right,black,xshift=0.0cm,yshift=-0.5cm] {$[\bm{x}^{\frac{1}{2}},\bm{x}^{1+\frac{1}{2}}\cdots \bm{x}^{{\rm{K}}-\frac{1}{2}}]$ }([yshift=0.0cm, xshift=-0.8cm]CNN2.north);
		\draw[arrow] ([yshift=0.0cm, xshift=0.0cm]INV2.north)--node [right,black,yshift=0.1cm] {$\bm{x}^{K-\frac{1}{2}}$ }(13.2cm, 1.4cm);		
		\end{tikzpicture}
	}
	\caption{Diagram of the proposed PFBS-(A)IR net for LDCT image reconstruction.}
	\label{fig:1}
\end{figure*}

\subsection{Implementation Details}
The best choice of step lengths $(\theta_0^1 \cdots \theta_{K-1}^1)$ and $(\theta^2_0,\cdots\theta^2_{K-1})$    with $\rm{K}$  iterations can be obtained by  end-to-end supervised training.
Let  $\{\bm{x}_{j},\bm{y}_{j}\}_{j=1}^J$ denote the training dataset with $J$ training samples, where $(\bm{x}_{j},\bm{y}_{j}) \in X\times Y$ denotes the pair of normal dose image and low dose projection data.
Then, the parameter $\theta$ is solved by the minimization  
\begin{equation}
\min_\theta \frac{1}{J}\sum_{j=1}^{J}L_\theta(\bm{x}_{j},\bm{y}_{j} ),
\label{loss}
\end{equation}
where the loss function is given as 
\begin{align}
L_\theta(\bm{x},\bm{y} ):=\| \mathcal{R}_\theta(\bm{y})-\bm{x}\|_X^2 ~~~{\rm{for}}~(\bm{x},\bm{y})\in X\times Y. \nonumber
\end{align}
To obtain the parameter $\theta$,  the back-propagation computations through all of the unrolled iterations are needed.

During the train process, we need to calculate gradients about $(\theta^1_0, \cdots, \theta^1_{K-1})$ and $(\theta^2_0,\cdots, \theta^2_{K-1})$, 
\begin{align}
\frac{\partial L_\theta}{\partial \theta^2_k}=\frac{\partial L_\theta}{\partial \bm{x}^K}
\cdot\frac{\partial \bm{x}^K}{\partial \bm{x}^{K-\frac{1}{2}}}\cdots \frac{\partial \bm{x}^{k+2}}{\partial \bm{x}^{k+\frac{3}{2}}}\cdot
\frac{\partial \bm{x}^{k+\frac{3}{2}}}{\partial \bm{x}^{k+1}}\cdot \frac{\partial \bm{x}^{k+1}}{\partial \theta^2_k},\\
\frac{\partial L_\theta}{\partial \theta^2_k}=\frac{\partial L_\theta}{\partial \bm{x}^K}
\cdot\frac{\partial \bm{x}^K}{\partial \bm{x}^{K-\frac{1}{2}}}\cdots \frac{\partial \bm{x}^{k+2}}{\partial \bm{x}^{k+\frac{3}{2}}}\cdot
\frac{\partial \bm{x}^{k+\frac{3}{2}}}{\partial \bm{x}^{k+1}}\cdot \frac{\partial \bm{x}^{k+1}}{\partial \theta^2_k},
\end{align}
where,
\begin{subequations}
	\label{backpro}
	\begin{numcases}
	{}\frac{\partial \bm{x}^{k+1}}{\partial\theta^2_k}=\frac{\partial{\rm{CNN}}([\bm{x}^{\frac{1}{2}},\bm{x}^{1+\frac{1}{2}}\cdots\bm{x}^{k+\frac{1}{2}}],\theta^2_k )}{\partial \theta^2_k},\\
	\frac{\partial \bm{x}^{k+\frac{1}{2}}}{\partial \theta^1_k}=\bm{A}^+(\bm{y}-\bm{A}\bm{x}^{k}),\\
	\frac{\partial \bm{x}^{k+\frac{1}{2}}}{\partial \bm{x}^{k}}=\bm{I}-\theta^1_k\bm{A}^+\bm{A},\\
	\frac{\partial \bm{x}^{k+1}}{\partial \bm{x}^{k+\frac{1}{2}}}=\frac{\partial{\rm{CNN}}([\bm{x}^{\frac{1}{2}},\bm{x}^{\frac{3}{2}}\cdots\bm{x}^{k+\frac{1}{2}}],\theta^2_k )}{\partial\bm{x}^{k+\frac{1}{2}}},\\
	\frac{\partial L_{\theta}}{\partial \bm{x}^K}=(\bm{x}^K-\bm{x}).	 
	\end{numcases}
\end{subequations}
After training the weights of the NN, we obtain an estimation of $\theta$.
For a low dose  input data $\bm{y}$,  the image can be reconstructed by applying  ${\rm{CNN}}([\bm{x}^{\frac{1}{2}},\bm{x}^{\frac{3}{2}}\cdots\bm{x}^{k+\frac{1}{2}}],\theta^2_k )$ and gradient descent, $\Lambda_{\theta^1_k}( \bm{x}^{k}, \bm{A}^+(\bm{A} \bm{x}^{k}-\bm{y}))$,  alternatively:
\begin{align}
\bm{y}&\rightarrow\bm{A}^+\bm{y}\rightarrow \cdots\rightarrow \bm{x}^{k}-\theta_k^1\bm{A}^+(\bm{Ax}^{k}-\bm{y})\nonumber\\
 &\rightarrow{\rm{CNN}}([\bm{x}^{\frac{1}{2}},\bm{x}^{1+\frac{1}{2}}\cdots\bm{x}^{k+\frac{1}{2}}],\theta^2_k ) \rightarrow \bm{x}^{k+1} \cdots \bm{x}^{\ast},\nonumber
\end{align} 
where $k=0,\cdots {\rm{K}-1}$ and $\bm{x}^{\ast}$ is the predicted image.

The training is performed with PyTorch \cite{paszke2017automatic} interface on a NVIDIA Titan GPU. 
Adam optimizer is used with the momentum parameter $\beta=0.9$, mini-batch size set to be $4$, and the learning rate set to be $10^{-4}$. 
At each stage, we use a standard CNN with the structure Conv$\rightarrow$BN$\rightarrow$ReLU, except the first block and the last block.
The BN layer is omitted for the first and  last block.  For all the Conv layers in the CNN, the kernel size is set as  $3\times3$.
The channel size is set to 64 and the outline of CNN is shown as Fig. \ref{fig:2}.
The model is trained with $50$ epochs.

\tikzstyle{Block} = [rectangle, minimum width=2cm, minimum height=.5cm, text centered, draw=blue, fill = blue!20]
\tikzstyle{Format} = [draw, thin, draw =white,fill=white]
\tikzstyle{FormatS} = [draw, thin, draw =white,fill=white,text width = 0.1cm]
\tikzstyle{cir} = [circle, text centered, draw=blue, fill = blue!20]
\begin{figure}[!htp]
	\centering
	\scalebox{0.68}{
		\begin{tikzpicture}[node distance=0.5cm]
		\path[->] node[FormatS] (Input){$\bm{x}^{\frac{1}{2}}$ $\bm{x}^{\frac{3}{2}}$ $\vdots$ $\bm{x}^{k+\frac{1}{2}}$};		
		\path[->] node[Block,right of=Input,rotate=-90,yshift=1.5cm] (layer1) {Conv};
		\path[->] node[Block,right of=layer1,rotate=-90] (layer2) {ReLU};
		\path[->] node[Block,right of=layer2,rotate=-90,yshift=1.0cm] (layer3) {Conv};
		\path[->] node[Block,right of=layer3,rotate=-90] (layer4) {BN};
		\path[->] node[Block,right of=layer4,rotate=-90] (layer5) {ReLU};
		\path[->] node[Block,right of=layer5,rotate=-90,yshift=1.0cm] (layer6) {Conv};
		\path[->] node[Block,right of=layer6,rotate=-90] (layer7) {BN};
		\path[->] node[Block,right of=layer7,rotate=-90] (layer8) {ReLU};
		\path[->] node[Block,right of=layer8,rotate=-90,yshift=1.0cm] (layer9) {Conv};
		\path[->] node[Block,right of=layer9,rotate=-90] (layer10) {ReLU};
		\path[->] node[cir,right of=layer10,xshift=1.0cm] (w1){$-$};
		\path[->] node[Format,right of=w1,xshift=1.0cm] (OutPut) {$\bm{x}^{k+1}$};
		\path[->] node[Format,below of=layer1,yshift=-1.0cm] () {$~~~~k\times 64$};
		\path[->] node[Format,below of=layer4,yshift=-1.0cm] () {$64\times 64$};
		\path[->] node[Format,below of=layer7,yshift=-1.0cm] () {$64\times 64$};
		\path[->] node[Format,below of=layer9,yshift=-1.0cm] () {$~~~~64\times1$};
		\draw [arrow] ([xshift=-2.5cm]Input) -- (layer1);
		\draw [arrow] (layer2) -- (layer3);
		\draw [fill] (5.5,0) circle [radius=.05];
		\draw [fill] (5.7,0) circle [radius=.05];
		\draw [fill] (5.9,0) circle [radius=.05];
		\draw [arrow] (layer8) -- (layer9);
		\draw [arrow] (layer10) -- ([ xshift=0.0cm]w1.west);
		\draw [arrow] (w1) -- ([ xshift=0.0cm]OutPut.west);
		\draw[arrow]([yshift=-1.0cm, xshift=0.8cm]Input.north)--++(0, 1.5cm)-|node [above,black,xshift=-5.0cm,yshift=0.0cm] {$\bm{x}^{k+\frac{1}{2}}$}([yshift=0.0cm, xshift=0.0cm]w1.north);
		\end{tikzpicture}
	}
	\caption{Diagram of CNN in Fig. \ref{fig:1}.}
	\label{fig:2}
\end{figure}
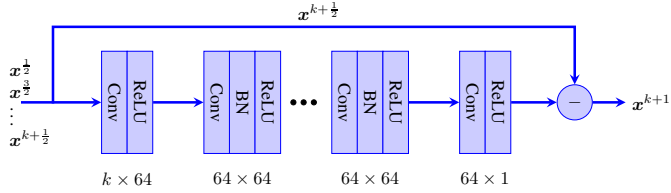

\section{Results}
\label{Experiment}
In this section, the proposed PFBS-IR and PFBS-AIR methods are evaluated using  prostate CT dataset, in comparison with TV-based IR method and a DL-based image postprocessing method, namely FBPConvNet. 

\subsection{Data}
To validate the performance of the proposed  methods at different dose levels, we simulated low dose projection data from their normal-dose counterparts. 
The normal dose dataset included 6400 normal-dose  prostate  CT  images of $256\times 256$ pixels per image from 100 anonymized scans.
The  LDCT projection data were simulated  by adding Poisson noise onto the normal-dose projection data \cite{ding2018statistical}:
\begin{eqnarray}
\bar{y}_i\sim {\rm{Poisson}}\{I_i\exp(-[\bm{A}\bm{\bm{x}}]_i)\}+ {\rm{Normal}}(0,\sigma_e^2) ,
\label{elemodel}
\end{eqnarray}
where $I_i$ is the incident X-ray intensity incorporating X-ray source illumination and the detector efficiency,
$\sigma_e^2$ is the background electronic
noise variance. The value of $\sigma_e^2$ was assumed to be stable for a
commercial CT scanner, and thus, the noise level was controlled by $I_i$

The simulated geometry for projection data include:  flat-panel  detector of $0.388~\rm{mm}\times 0.388~\rm{mm}$ pixel size, 
$600$ projection views  evenly  spanning a $360^{\circ}$ circular orbit,  $512$ detector bins for each projection, 
$100.0~\rm{cm}$ source to detector distance and $50.0~\rm{cm}$  source to isocenter distance.
In  the simulation, the noise level is controlled by X-ray intensity $I_i$, which is set uniformly, \emph{i.e.} , $I_1=I_i, i=1,\cdots N_d$.
The noise level was set to be uniform, \emph{i.e.}, $I_i=10^5$, $5\times10^4$, $10^4,5\times10^3$ respectively. 
Then, the projection data for reconstruction were obtained by taking logarithm on projection data $\bar{\bm{y}}$.
$80$ scans were included in the training set, and the rest $20$ scans were included in the testing set.

\subsection{Methods for comparison}
The performance of the proposed methods is evaluated in comparsion with  FBP (an AR method), TV (an IR method) and FBPConvNet (a DL-based image postprocessing method).
\subsubsection{TV-based IR method}
 The TV-based IR method was solved by ADMM:
\begin{eqnarray}
\nonumber
\left\{\begin{array}{ccc}
\bm{x}^{k+1}&=&\arg\min_{\bm{x}}\frac{1}{2}\|\bm{Ax}-\bm{y}\|_2^2+\frac{\mu}{2}\|\nabla\bm{x}-\bm{z}^{k}+\frac{\bm{p}^k}{\mu}\|_2^2,\\
\bm{z}^{k+1}&=&\arg\min_{\bm{z}}\lambda\|\bm{z}\|_1+\frac{\mu}{2}\|\bm{z}-(\nabla\bm{x}^{k+1}+\frac{\bm{p}^k}{\mu})\|_2^2,\\
\bm{p}^{k+1}&=&\bm{p}^{k}+\mu(\nabla\bm{x}^{k+1}-\bm{z}^{k+1} ),
\end{array}\right.
\end{eqnarray}
where $\bm{z}$ is the auxiliary variable,   $\bm{p}$ is the dual variable, $\mu$ is the algorithm parameter and  $\nabla$ is the gradient operator.
The parameters $\lambda,\mu$ of the TV-based IR method were manually optimized.
Specifically, the regularization parameter $\lambda$ for the TV-based IR method was set to $0.01$ for $I_i=10^5$ and $I_i=5\times10^4$, $0.03$ for $I_i=10^4$ and  $0.05$ for $I_i=5\times10^3$,  which yielded the best performance.
\subsubsection{FBPConvNet}
FBPConvNet \cite{jin2017deep} is a state-of-the-art DL technique, in which a residual CNN with U-net architecture is trained to directly denoise the FBP. It has been shown to outperform other DL-based  methods for CT reconstruction. 
\subsection{Results}

For our methods, we set  $\rm{K}=10$. 
For every stage, 5-block modified CNN is applied. 
For PFBS-AIR, $\bm{A}^+$ is set to be the FBP operator, while for PFBS-IR, $\bm{A}^+=\bm{A}^T$.

The three metrics, peak signal to noise ratio (PSNR), root mean square error (RMSE)  and structural similarity index measure (SSIM) \cite{wang2004image}, are chosen for quantitative evaluation of image quality.
PSNR is defined as 
\begin{equation}
{\rm{PSNR}}(\bm{x},\bm{x}^\ast)=10\log_{10}\left(\frac{\max(\bm{x}.\ast \bm{x})}{\|\bm{x}-\bm{x}^\ast\|_2^2}\right),
\end{equation}
where $.\ast$ denotes element-wise multiplication, $\bm{x}^\ast$ is the reconstructed image and $\bm{x}$ is the ground truth (normal dose image).
RMSE is defined as
\begin{equation}
{\rm{RMSE}}=\sqrt{\frac{\sum_{i=1}^N(x^\ast_i-x_i)^2}{N}},
\end{equation}
where $N$ is the number of pixels and $i$ is the pixel index.

The quantitative results for the reconstructed images are given in Table \ref{Comparision}.
Table \ref{Comparision} shows the means and standard deviations (STD) of PSNR, RMSE and SSIM for all the images reconstructed  with different low dose levels.
The table suggests that our method achieved superior performance for all low-dose levels.
TV had larger PSNRs, smaller RMSEs and larger SSIMs than FBP method as expected.
The  DL-based methods improved the reconstructed results from FBP and TV, among which PFBS-AIR had the best reconstruction quality in terms of PSNR, RMSE and SSIM.
\begin{table*}[htb]
	\caption{Quantitative reconstruction results for all images.}
	\scalebox{1}{
		\begin{tabular}{p{2.1cm}p{0.8cm}p{2.3cm}p{2.3cm}p{2.3cm}p{2.3cm}p{2.3cm}p{3.3cm}}
			\hline     
			\hline   
			{Dose level}                    &            & FBP                   & TV                       & FBPConvNet              & PFBS-IR             & PFBS-AIR          \\
			\hline
			\multirow{3}{*}{$I_i=10^5$}     &PSNR        &   $41.6739\pm1.4145$  &     $44.9089\pm1.4348$   &    $47.0168\pm1.4717$   & $45.9053\pm1.4028$  &    $\bold{50.1927\pm1.7112}$          \\
			                                &RMSE        &   $~~0.0033\pm0.0002$ &     $~~0.0023\pm0.0002$  &    $~~0.0018\pm0.0002$  & $1.4028\pm0.0021$   &    $\bold{~~0.0013\pm0.0002}$          \\
			                                &SSIM        &   $~~0.9933\pm0.0008$ &     $~~0.9971\pm0.0007 $ &    $~~0.9975\pm0.0007$  & $ 0.9974\pm0.0007$  &    $\bold{~~0.9986\pm0.0005}$          \\
			\hline
			\multirow{3}{*}{$I_i=5\times10^4$}&PSNR      &   $40.5805\pm1.4185$  &     $43.5493 \pm1.4465$  &   $45.9569\pm1.4759$    &$43.9504\pm1.4784$   &    $\bold{49.2162\pm1.7964}$          \\
			                                &RMSE        &   $~~0.0038\pm0.0003$ &     $~~0.0027\pm0.0003$  &    $~~0.0021\pm0.0003$  &$0.0026\pm0.0002$    &    $\bold{~~0.0014\pm0.0002}$          \\
			                                &SSIM        &   $~~0.9902\pm0.0017$ &     $~~0.9954\pm0.0013 $ &    $~~0.9965\pm0.0009$  &$0.9947\pm0.0008$    &    $\bold{~~0.9983\pm0.0007}$          \\
			\hline
			\multirow{3}{*}{$I_i=10^4$}     &PSNR        &   $35.9736\pm1.5382$  &     $39.8860\pm1.5873$   &    $43.2968\pm1.5592$   &$42.6190\pm1.4974$   &    $\bold{45.7214\pm1.7246}$          \\
			                                &RMSE        &   $~~0.0066\pm0.0008$ &     $~~0.0041\pm0.0005$  &    $~~0.0028\pm0.0003$  &$0.0031\pm0.0003$    &    $\bold{~~0.0021\pm0.0003}$          \\
			                                &SSIM        &   $~~0.9636\pm0.0093$ &     $~~0.9881\pm0.0039$  &    $~~0.9942\pm0.0014$  &$0.9925\pm 0.0013$   &    $\bold{~~0.9964\pm0.0011}$          \\
			\hline
			\multirow{3}{*}{$I_i=5\times10^3$}&PSNR      &   $33.2052\pm1.5943$  &     $38.2131\pm1.6229$   &    $41.6542\pm1.5140$   &$41.9751\pm1.5488$   &    $\bold{44.0442\pm1.7250}$          \\
			                                &RMSE        &   $~~0.0092\pm0.0012$ &     $~~0.0050\pm0.0007$  &    $~~0.0034\pm0.0004$  &$0.0033\pm0.0004$    &    $\bold{~~0.0026\pm0.0004}$          \\
		                                	&SSIM        &   $~~0.9288\pm0.0187$ &     $~~0.9827\pm0.0059$  &    $~~0.9922\pm0.0017$  &$0.9923\pm0.0018$    &    $\bold{~~0.9951\pm0.0015}$          \\
			\hline                                                                                                                       
			\hline
		\end{tabular}        
	}
	\label{Comparision}
\end{table*}

\begin{figure}[htb]
	\begin{center}
		\begin{tabular}{c@{\hspace{2pt}}c@{\hspace{2pt}}c@{\hspace{2pt}}c@{\hspace{2pt}}c@{\hspace{2pt}}c@{\hspace{2pt}}c@{\hspace{2pt}}c@{\hspace{2pt}}c@{\hspace{2pt}}c}
			\includegraphics[width=.45\linewidth,height=.3\linewidth]{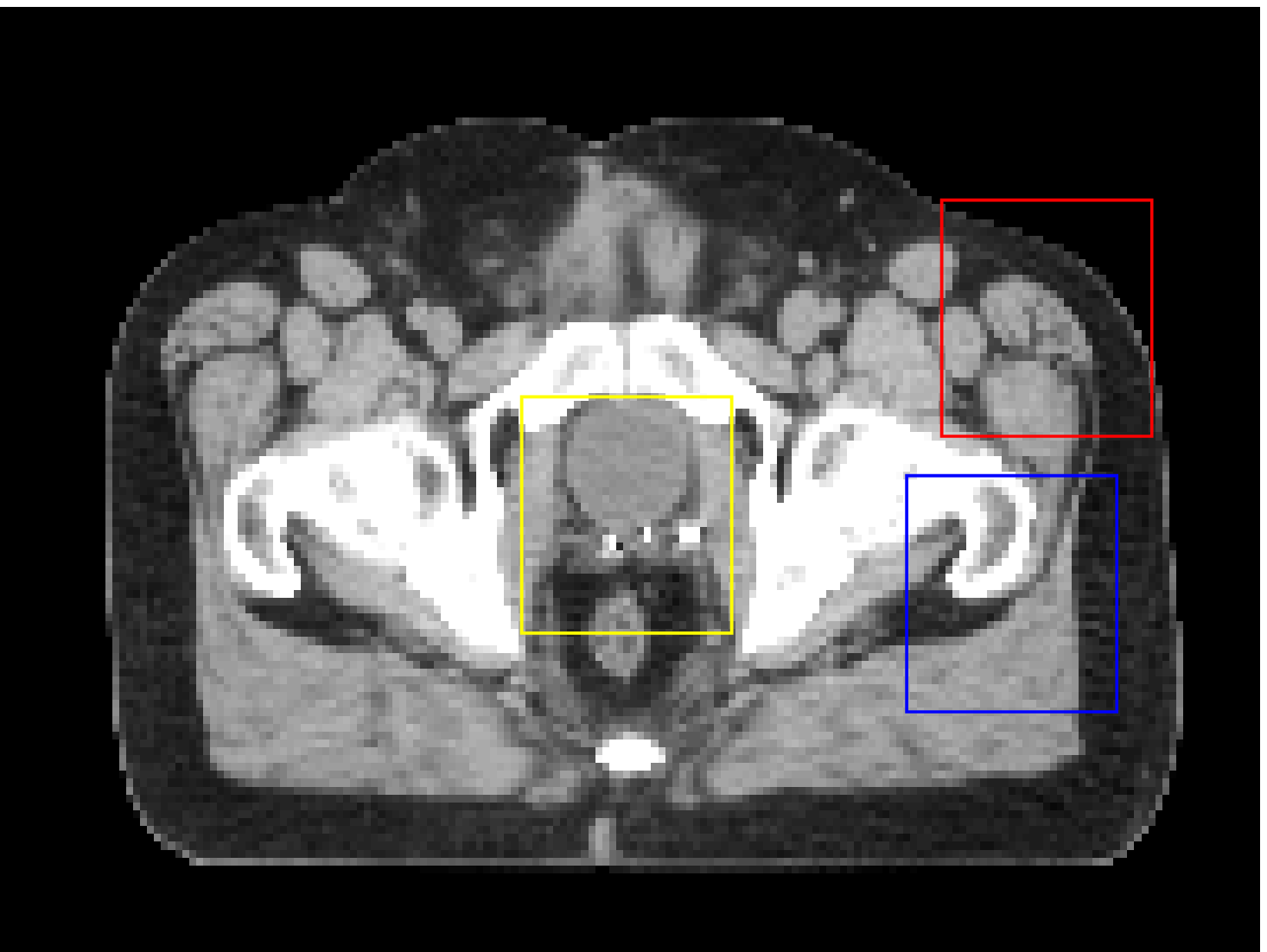}&
			\includegraphics[width=.45\linewidth,height=.3\linewidth]{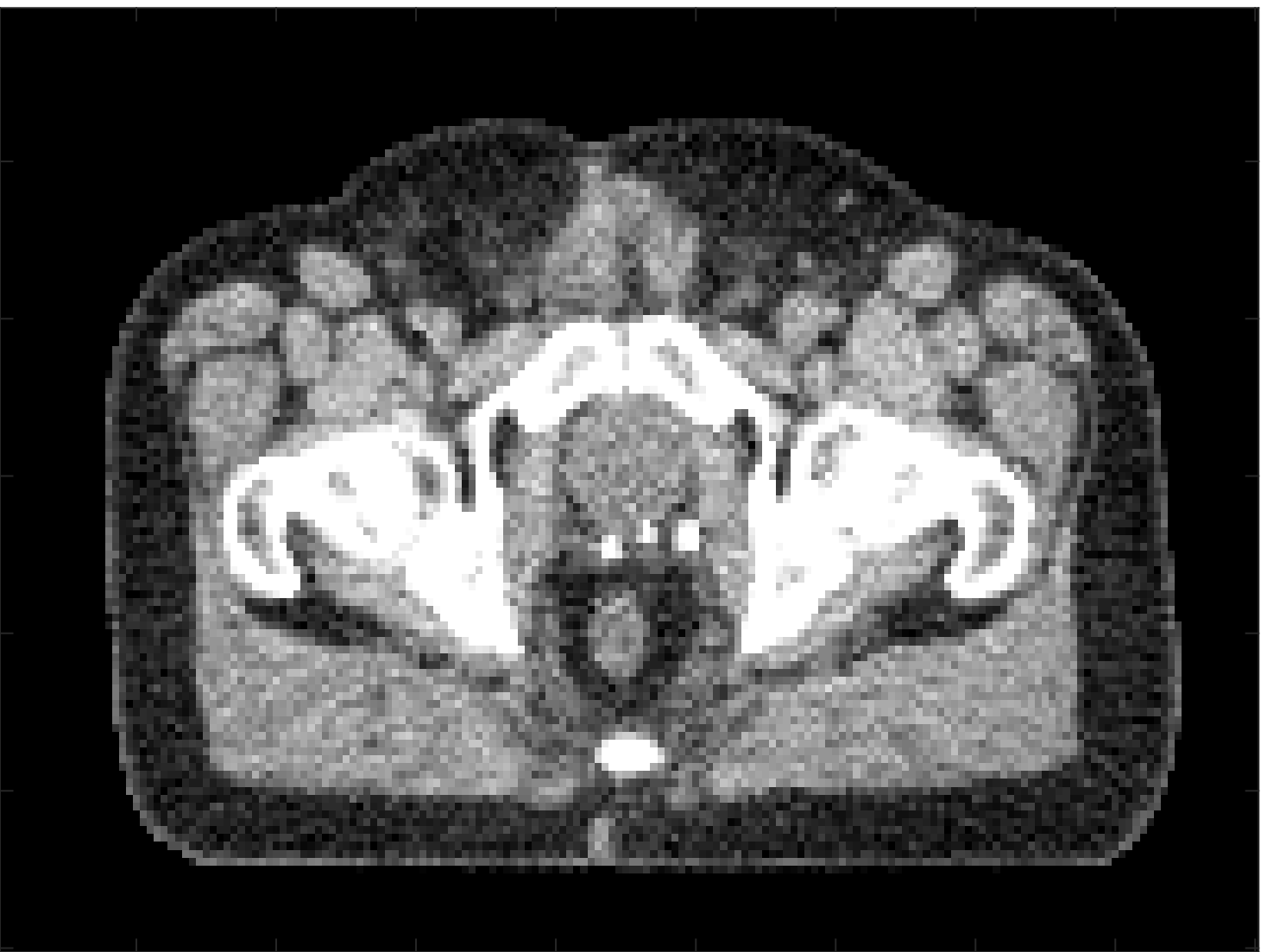}\\
            NDCT&
			FBP\\
			\includegraphics[width=.45\linewidth,height=.3\linewidth]{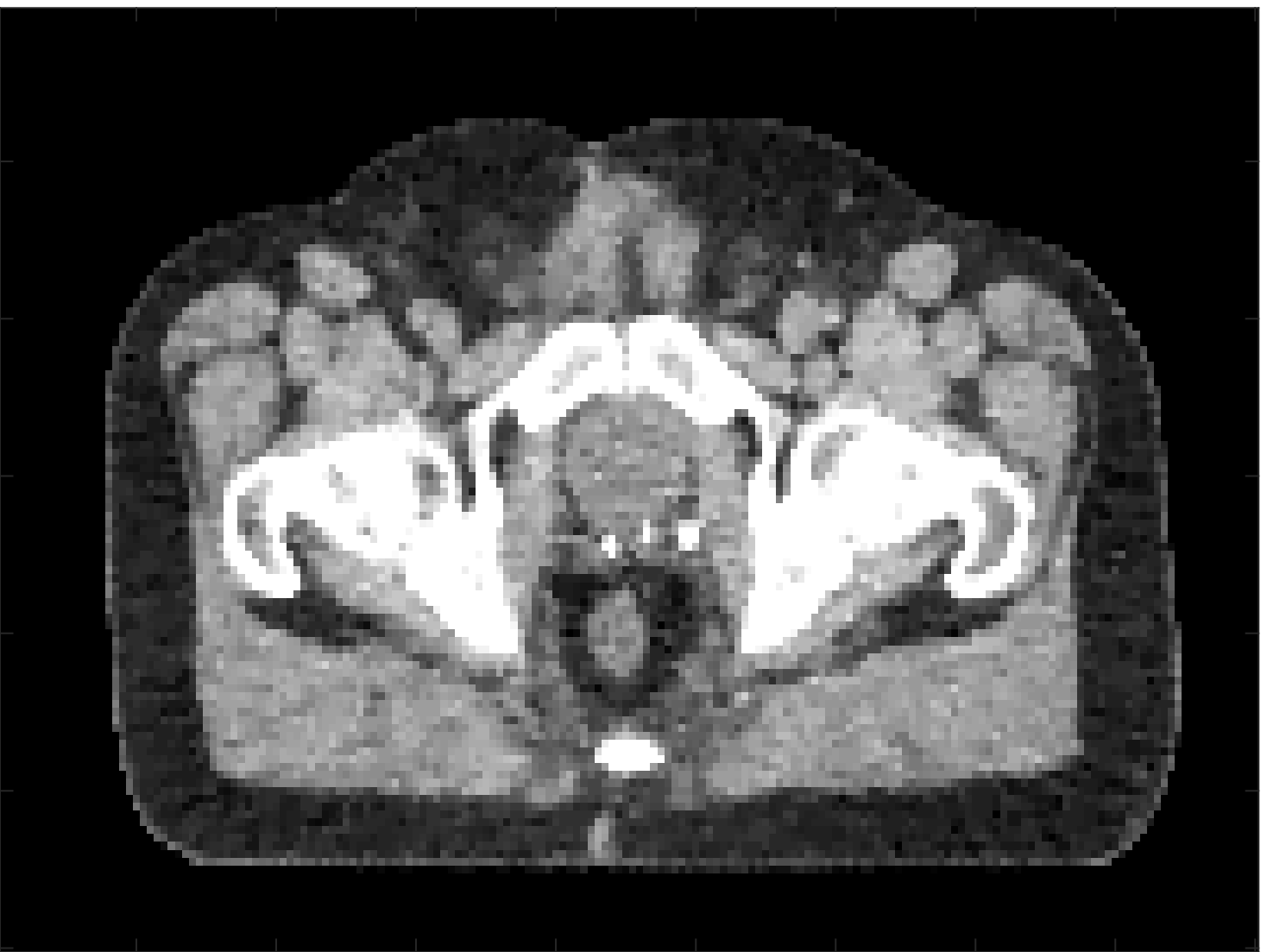}&
			\includegraphics[width=.45\linewidth,height=.3\linewidth]{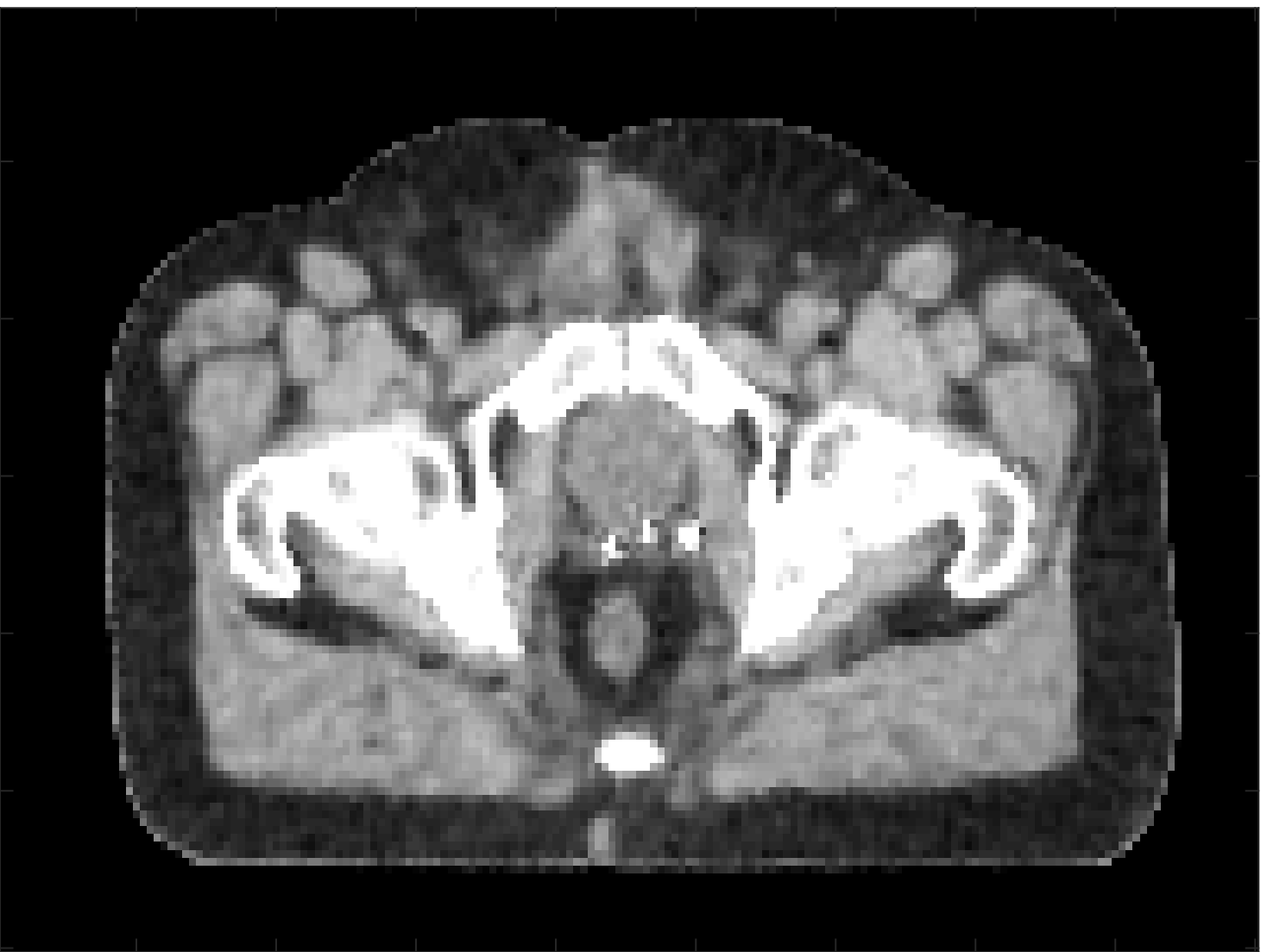}\\
				TV&
				{FBPConvNet}\\
			\includegraphics[width=.45\linewidth,height=.3\linewidth]{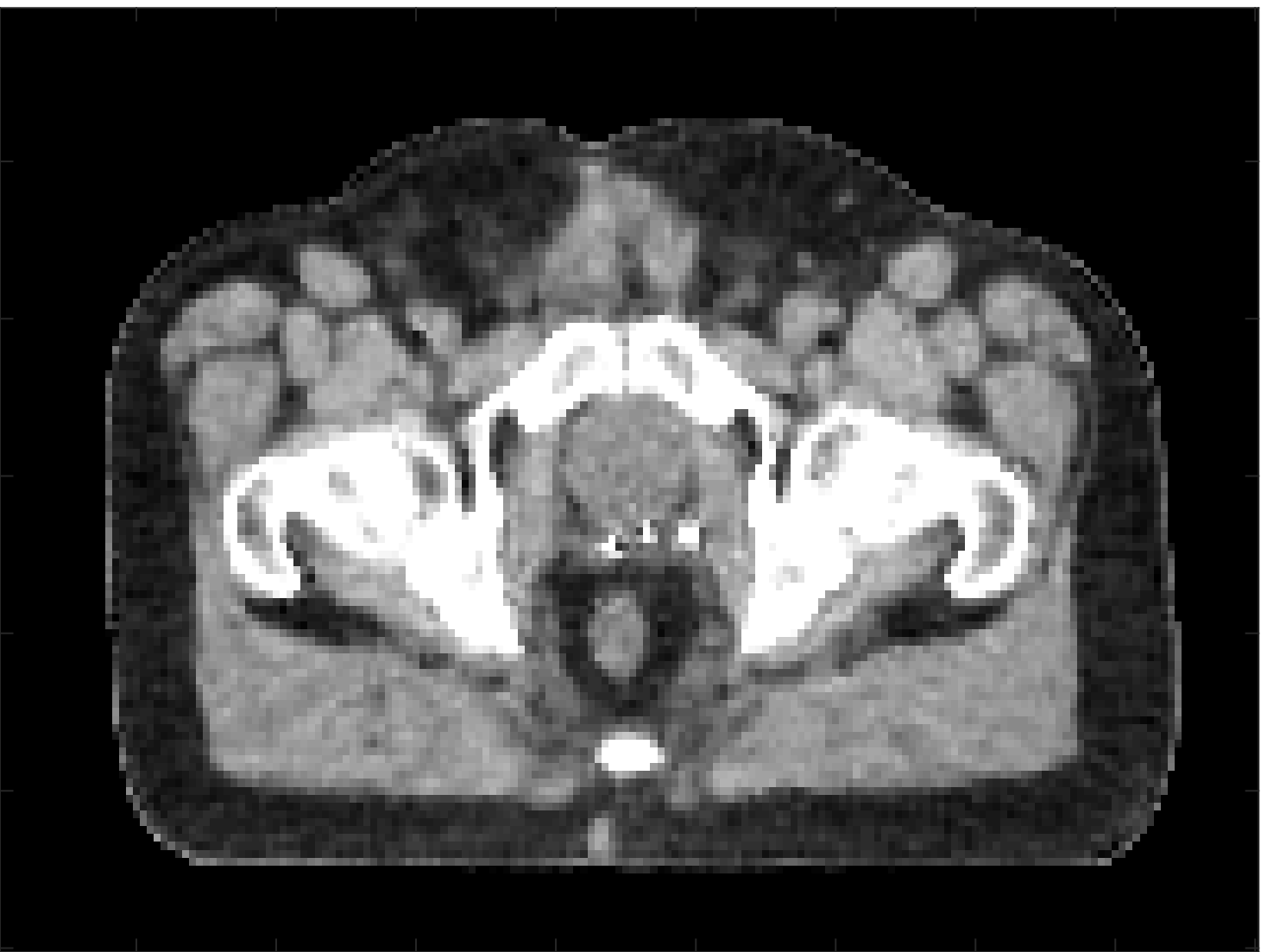}&
			\includegraphics[width=.45\linewidth,height=.3\linewidth]{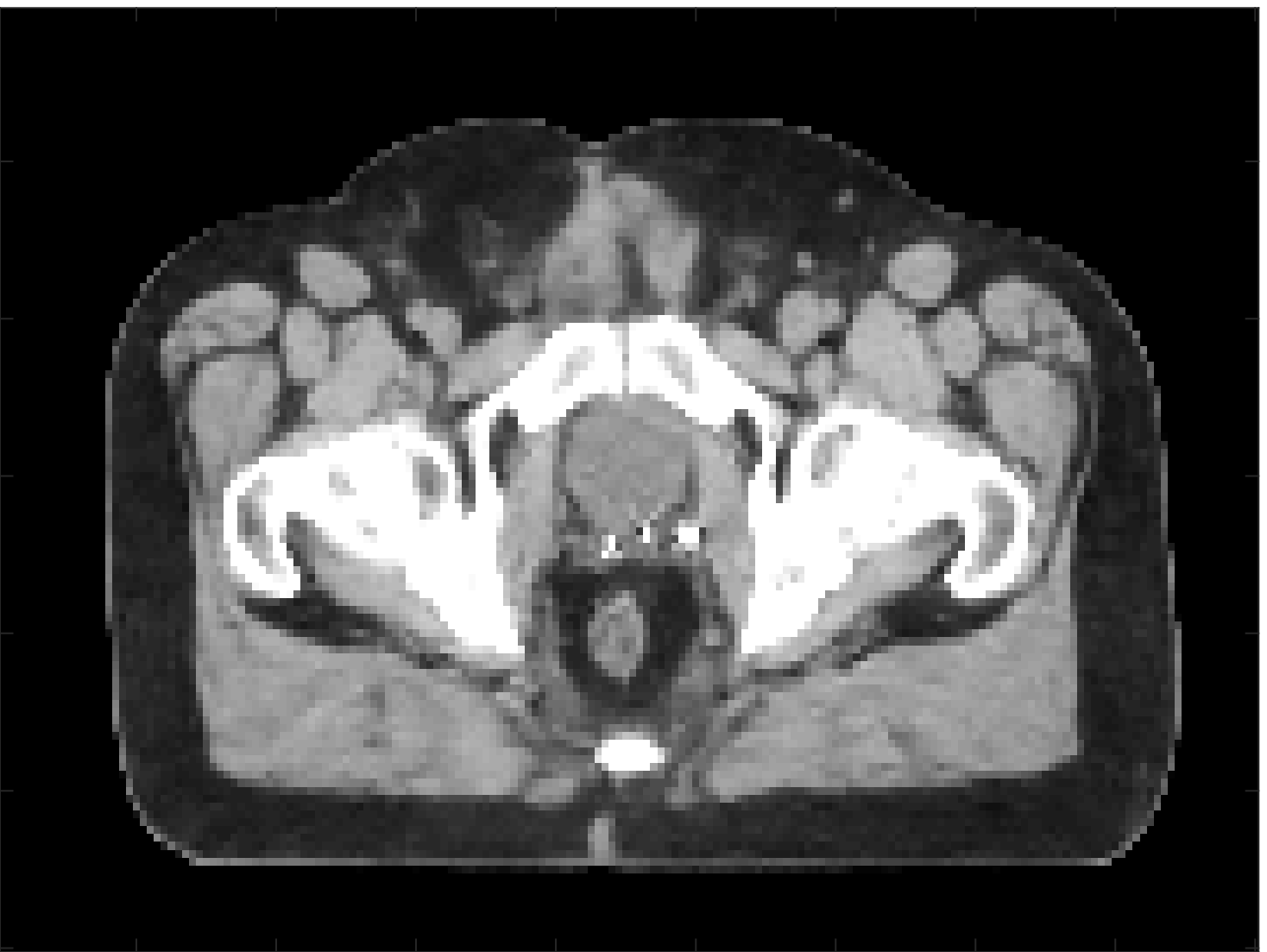}\\	
			PFBS-IR&
			PFBS-AIR
		\end{tabular}
		\caption {Reconstruction results at dose level  $I_i=5\times10^4$.}
		\label{slice2rec_36_50000}
	\end{center}
\end{figure}

A representative slice from all methods is showed in Fig. \ref{slice2rec_36_50000} with the dose level $I_i=5\times10^4$. The displayed window is set to $[- 150,150]$HU for all Figures. 
And their  zoomed-in images are presented in Fig. \ref{slice2rec_36_50000} 
indicated by the arrows in Fig. \ref{slice2rec_36_50000},  while FBPConvNet and PFBS-IR method were blurred,  PFBS-AIR had superior reconstruction quality.

\begin{figure}[htb]
		\begin{tabular}{c@{\hspace{2pt}}}
			\includegraphics[width=.96\linewidth,height=.16\linewidth]{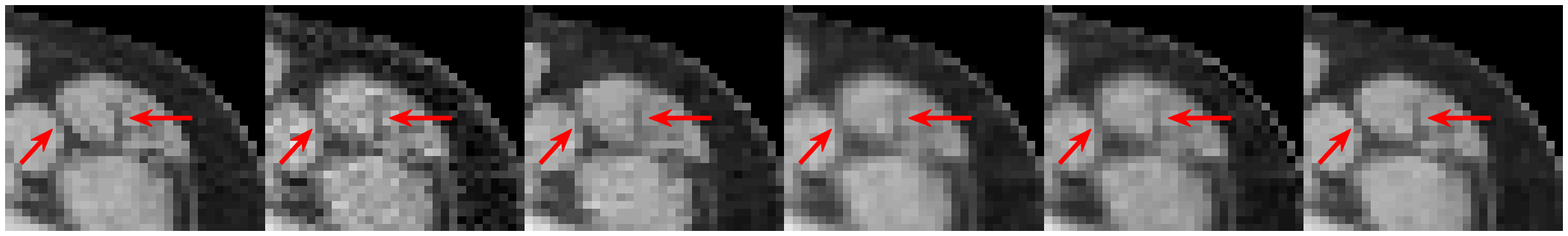}\\
			\includegraphics[width=.96\linewidth,height=.16\linewidth]{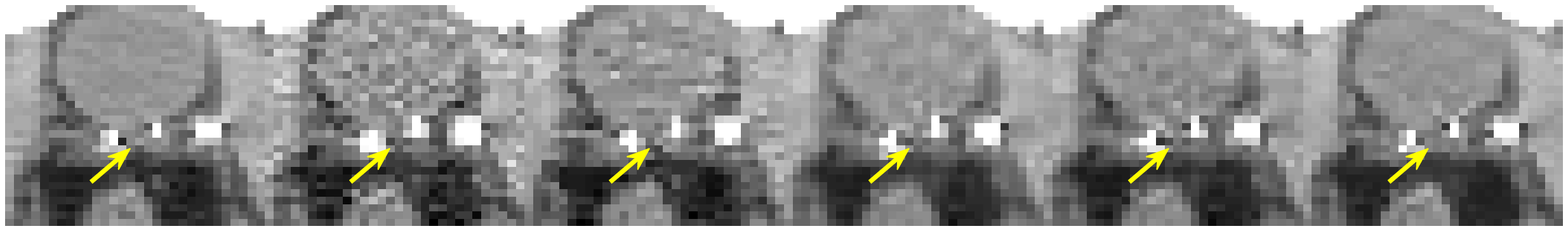}\\
			\includegraphics[width=.96\linewidth,height=.16\linewidth]{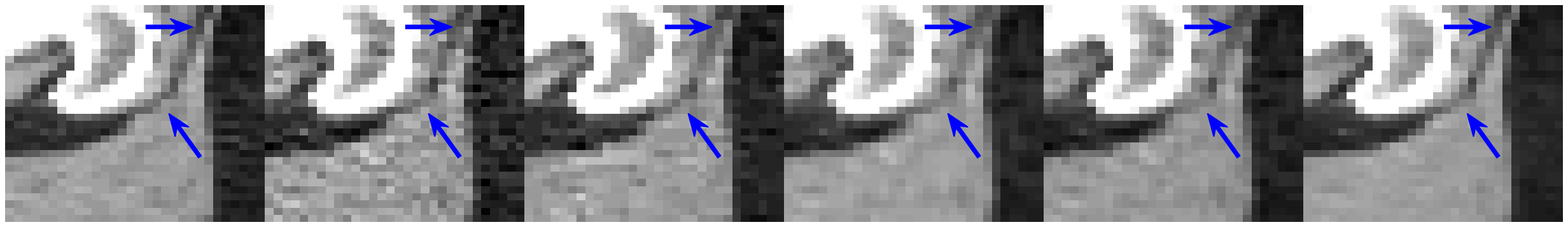}\\
		\footnotesize{~~~~NDCT~~~~~~FBP~~~~~~~~~~~TV~~~~~FBPConvNet~PFBS-IR~~~PFBS-AIR~~}
		\end{tabular}
		\caption{Zoom-in reconstruction results at dose level $I_i=5\times10^4$.  Three rows from up to bottom correspond to the red, yellow and blue  boxes in Figure \ref{slice2rec_36_50000} respectively, with differences highlighted in arrows.}
		\label{Zoomslice2rec_36_50000}
\end{figure}

\begin{figure}[htb]
	\begin{center}
		\begin{tabular}{c@{\hspace{2pt}}c@{\hspace{2pt}}c@{\hspace{2pt}}c@{\hspace{2pt}}c@{\hspace{2pt}}c@{\hspace{2pt}}c@{\hspace{2pt}}c@{\hspace{2pt}}c@{\hspace{2pt}}c}
			\includegraphics[width=.45\linewidth,height=.3\linewidth]{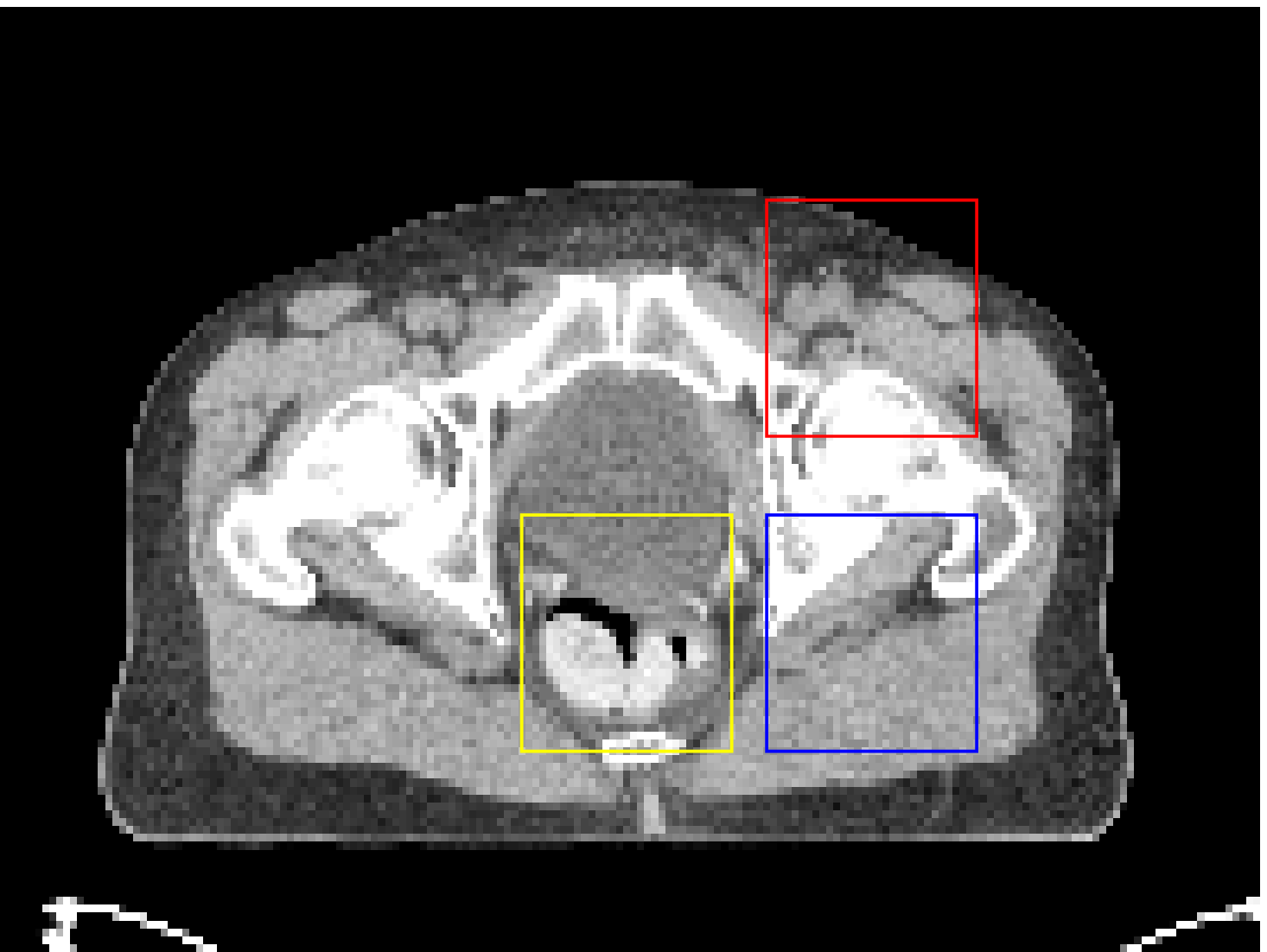}&
			\includegraphics[width=.45\linewidth,height=.3\linewidth]{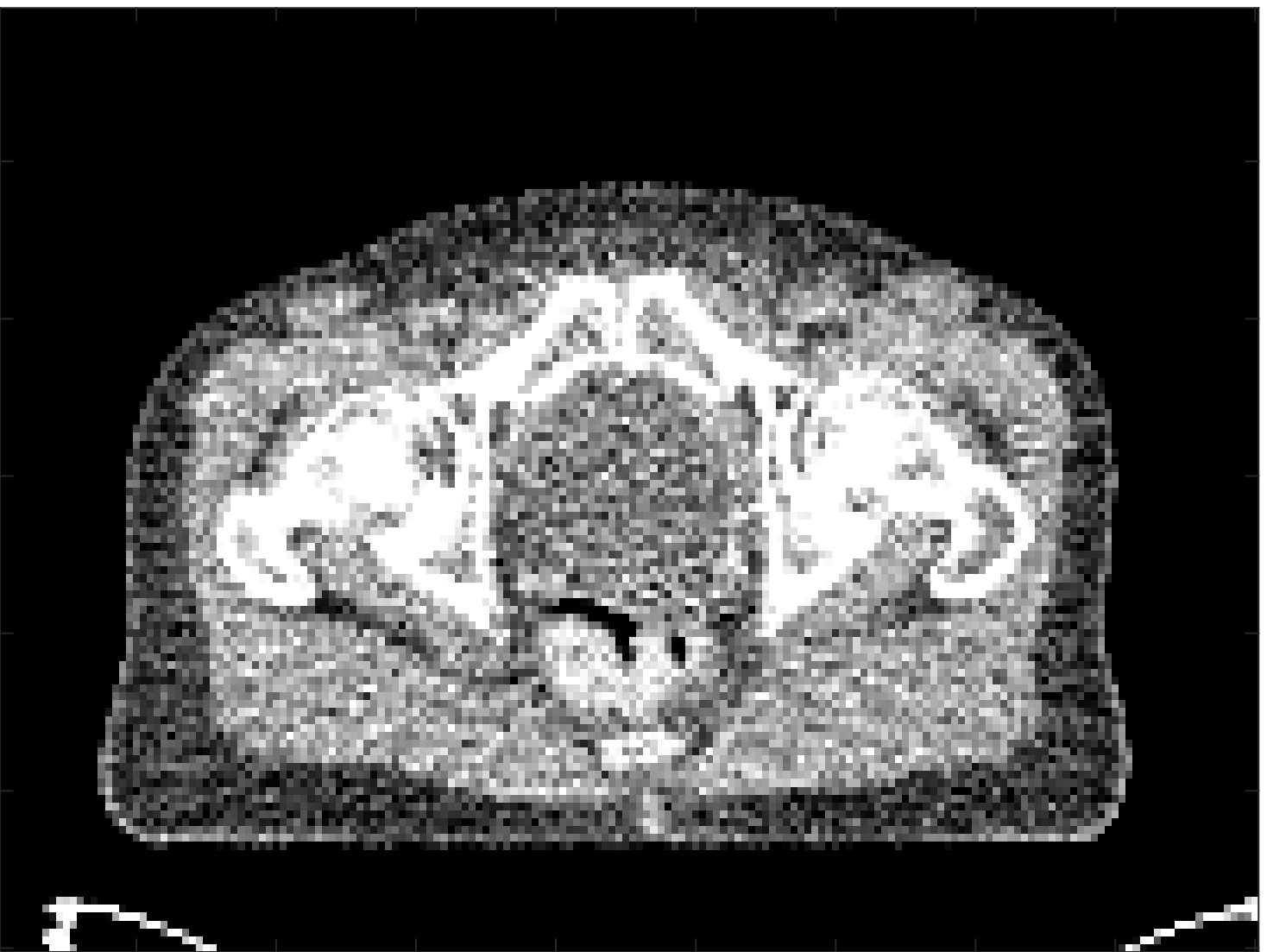}\\
			NDCT&
			FBP\\
			\includegraphics[width=.45\linewidth,height=.3\linewidth]{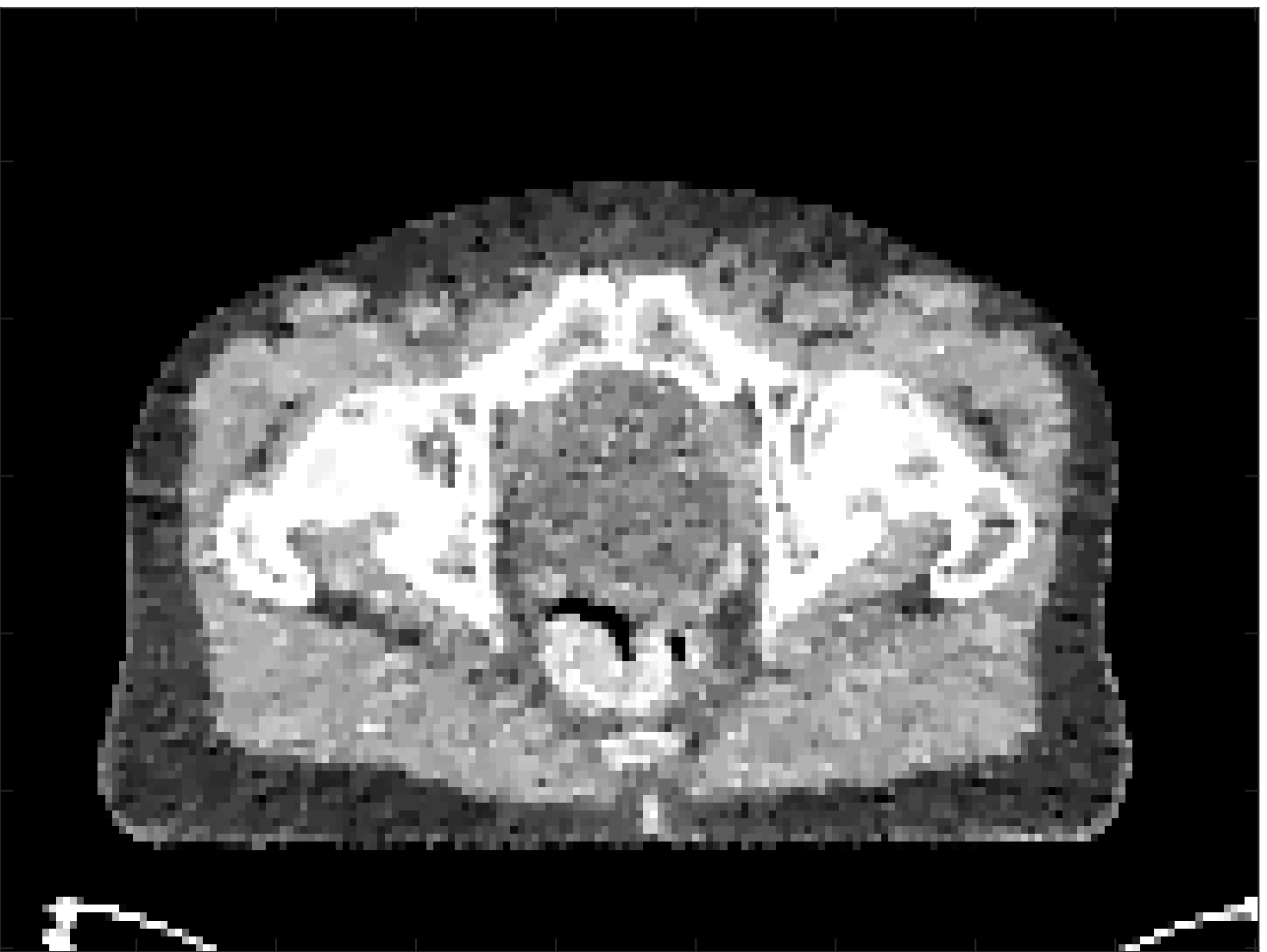}&
			\includegraphics[width=.45\linewidth,height=.3\linewidth]{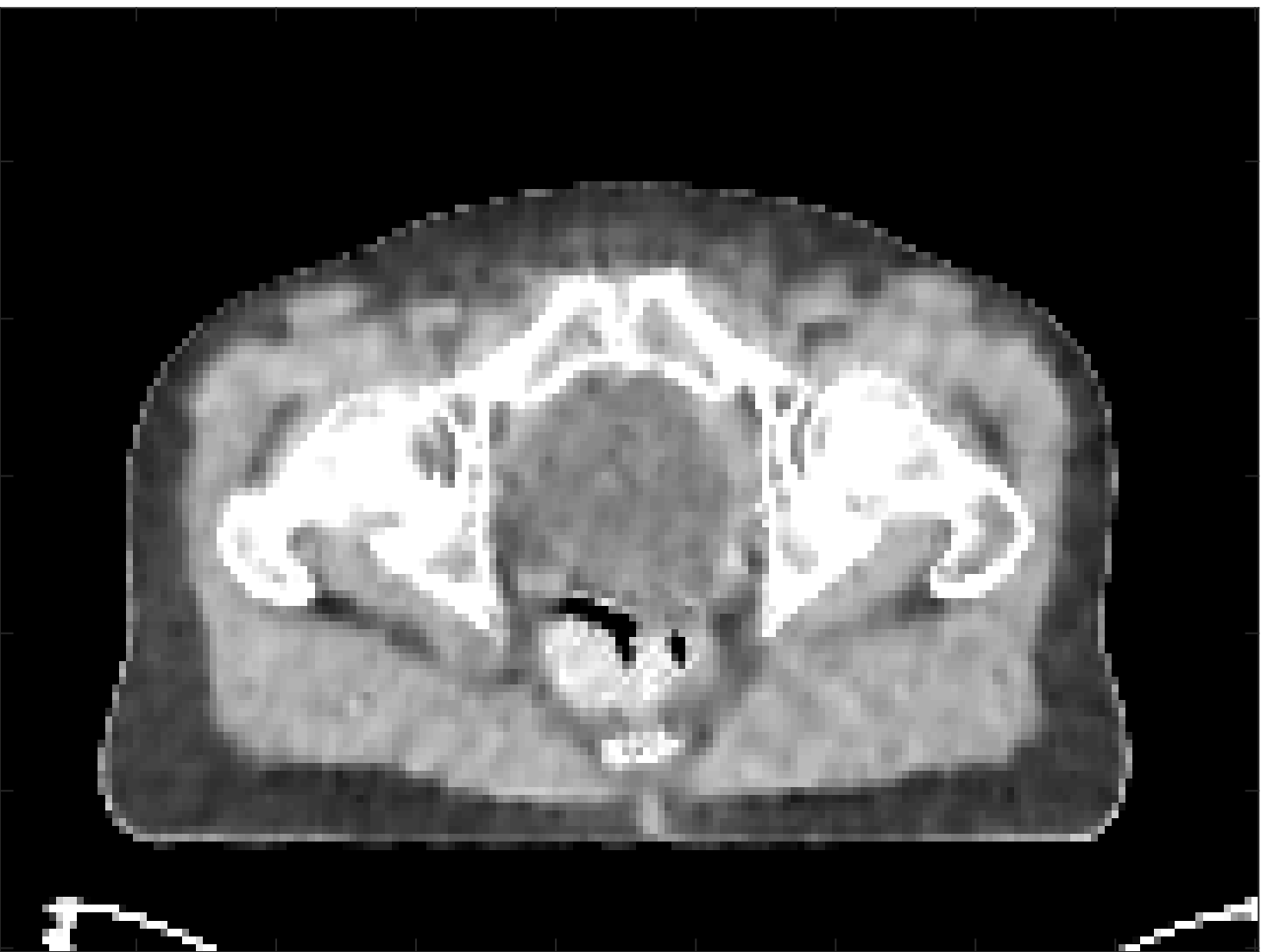}\\
			TV&
			{FBPConvNet}\\
			\includegraphics[width=.45\linewidth,height=.3\linewidth]{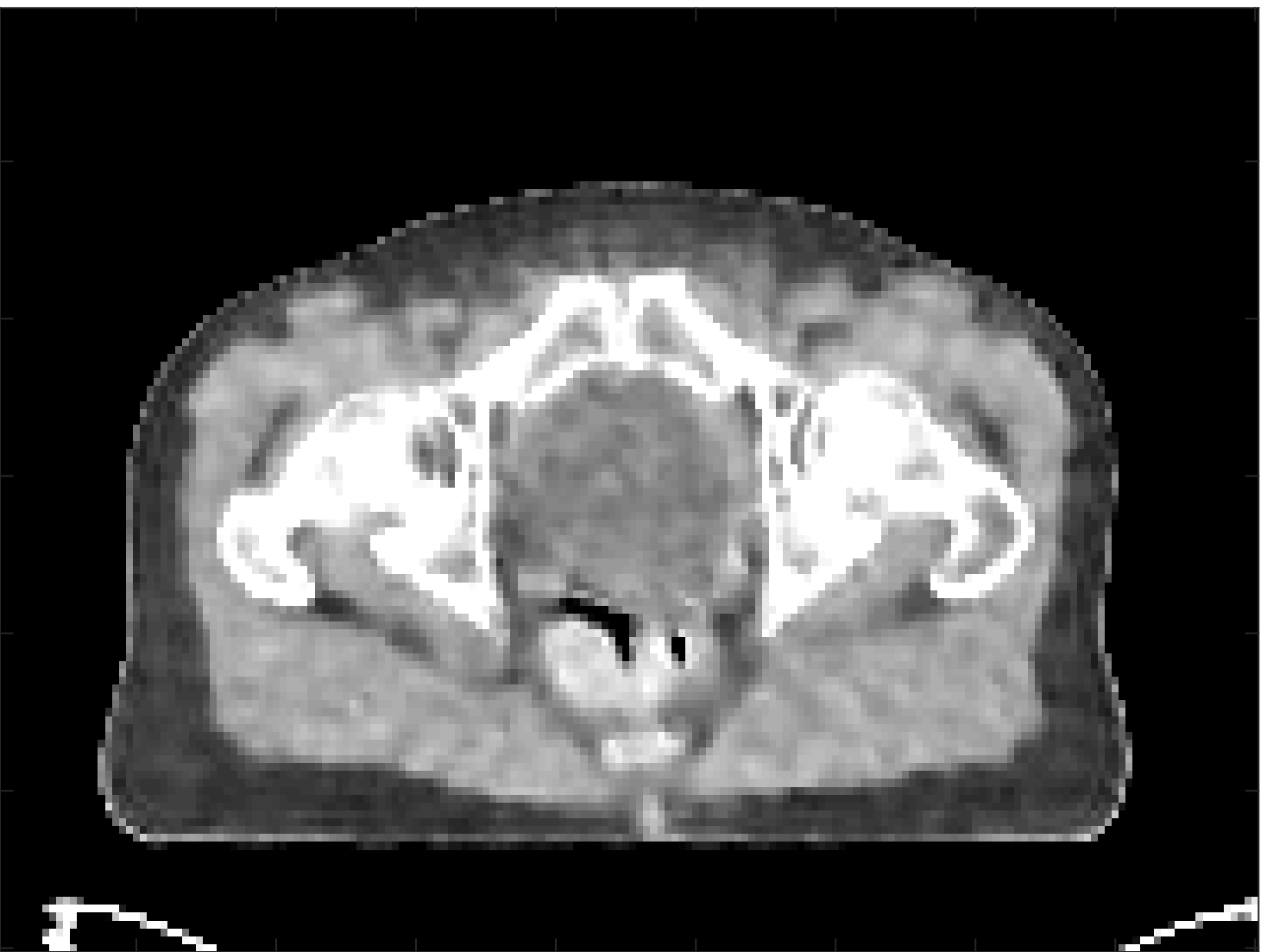}&
			\includegraphics[width=.45\linewidth,height=.3\linewidth]{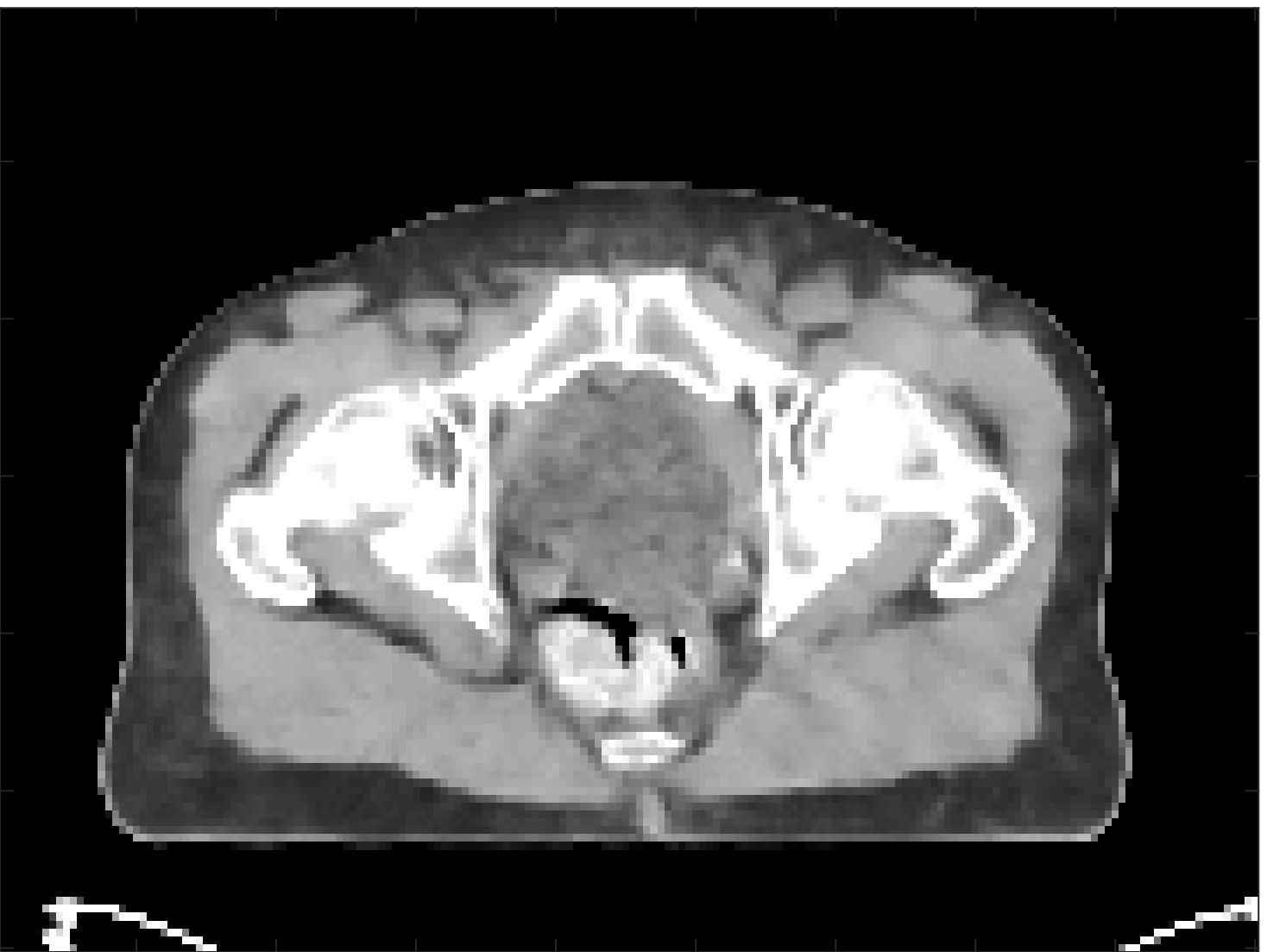}\\	
			PFBS-IR&
			PFBS-AIR
		\end{tabular}
		\caption {Reconstruction results at dose level  $I_i=10^4$.}
		\label{slice12rec_38_10000}
	\end{center}
\end{figure}

\begin{figure}[htb]
	\begin{center}
		\begin{tabular}{c@{\hspace{2pt}}}
			\includegraphics[width=.96\linewidth,height=.16\linewidth]{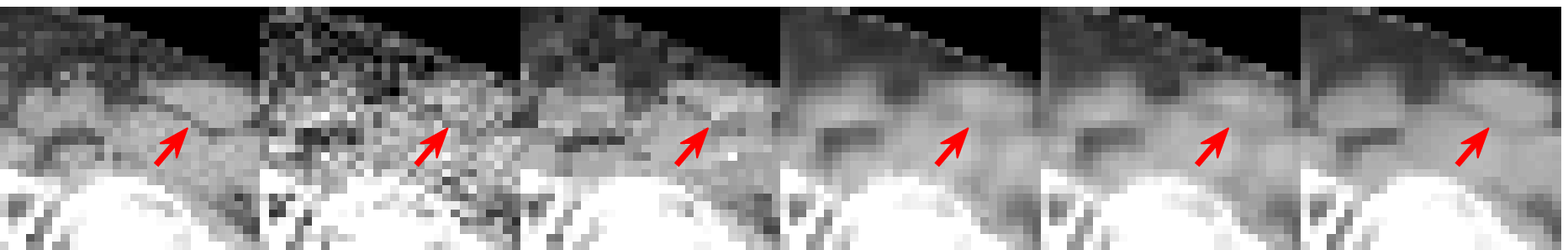}\\
			\includegraphics[width=.96\linewidth,height=.16\linewidth]{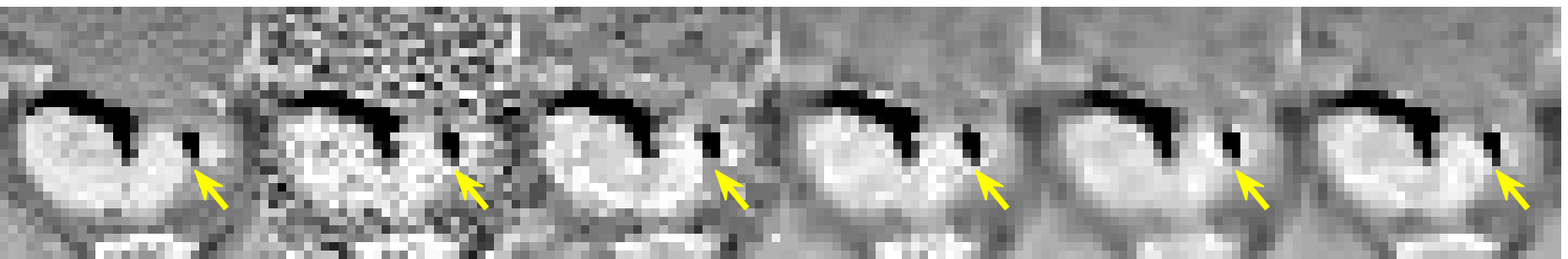}\\
			\includegraphics[width=.96\linewidth,height=.16\linewidth]{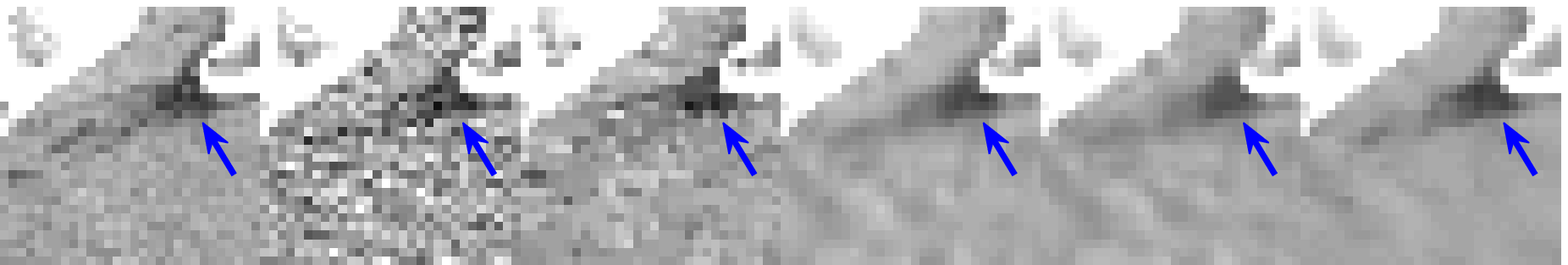}\\
		\footnotesize{~~~~NDCT~~~~~~FBP~~~~~~~~~~~TV~~~~~FBPConvNet~PFBS-IR~~~PFBS-AIR~~}
		\end{tabular}
		\caption{Zoom-in reconstruction results at dose level  $I_i=10^4$.  Three rows from up to bottom correspond to the red, yellow and blue  boxes in Fig.  \ref{slice12rec_38_10000} respectively, with differences highlighted in arrows.}
		\label{Zoomslice12rec_38_10000}
	\end{center}
\end{figure}

\begin{figure}[htb]
	\begin{center}
		\begin{tabular}{c@{\hspace{2pt}}c@{\hspace{2pt}}c@{\hspace{2pt}}c@{\hspace{2pt}}c@{\hspace{2pt}}c@{\hspace{2pt}}c@{\hspace{2pt}}c@{\hspace{2pt}}c@{\hspace{2pt}}c}
			\includegraphics[width=.45\linewidth,height=.3\linewidth]{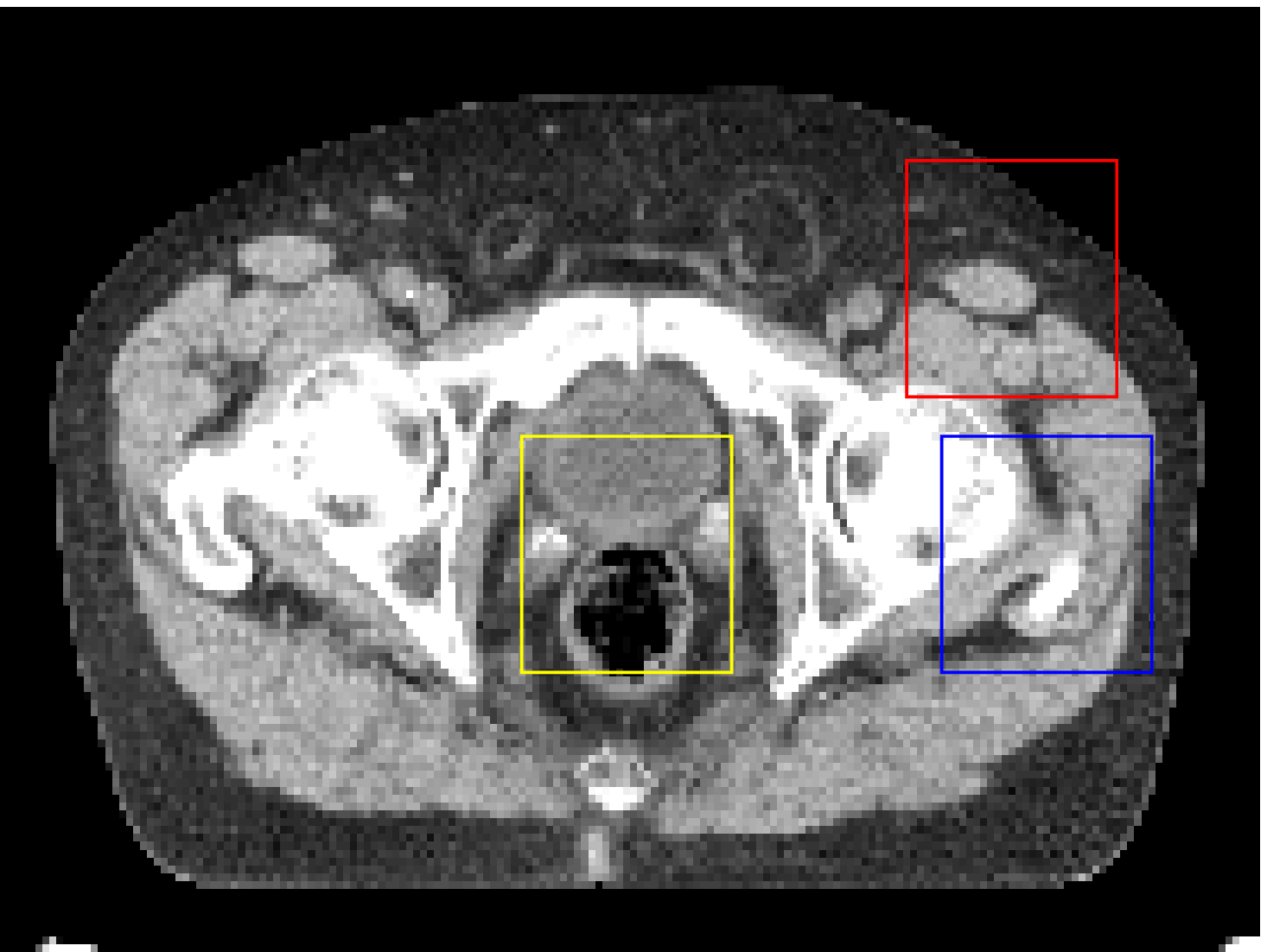}&
			\includegraphics[width=.45\linewidth,height=.3\linewidth]{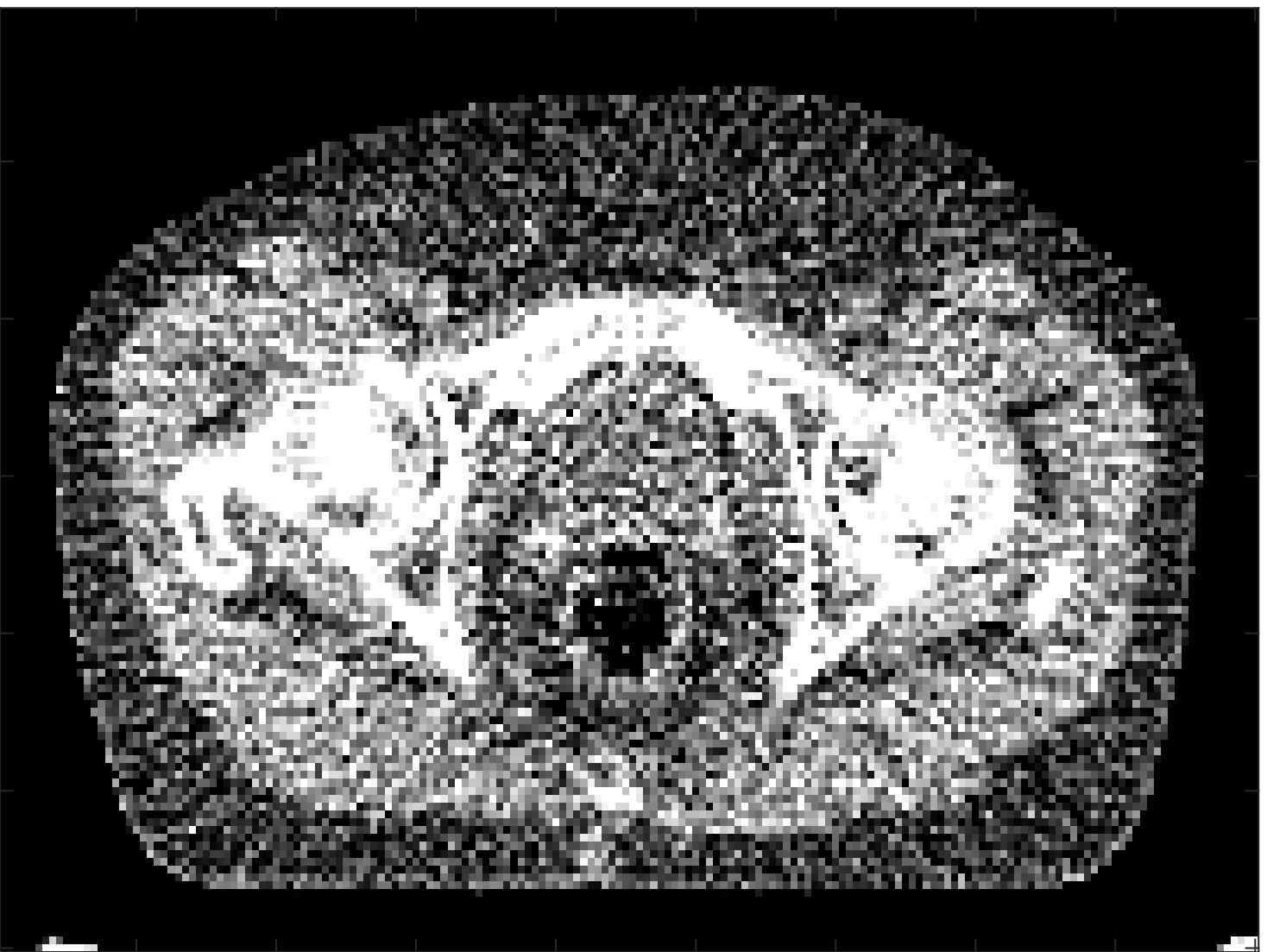}\\
			NDCT&
			FBP\\
			\includegraphics[width=.45\linewidth,height=.3\linewidth]{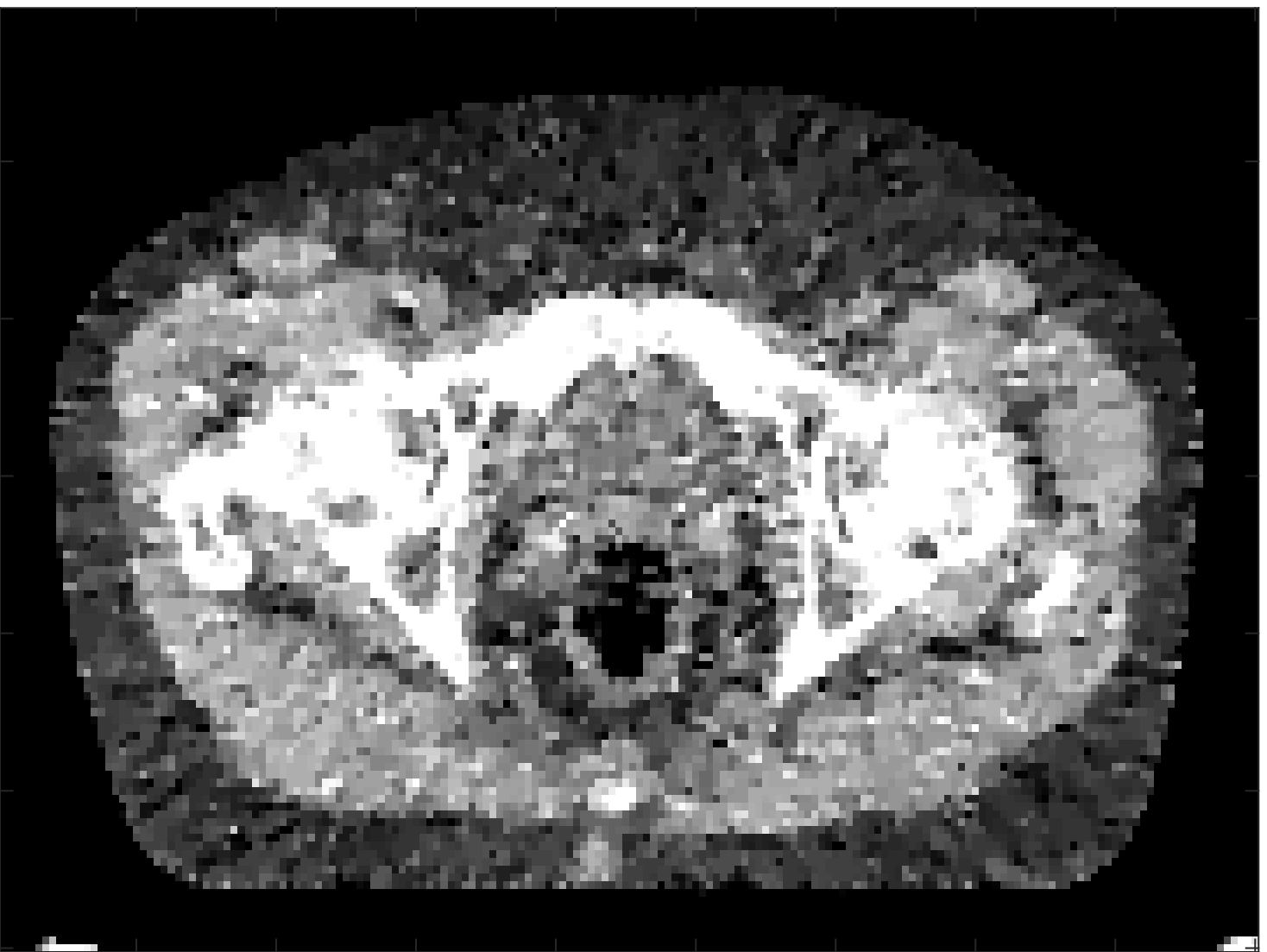}&
			\includegraphics[width=.45\linewidth,height=.3\linewidth]{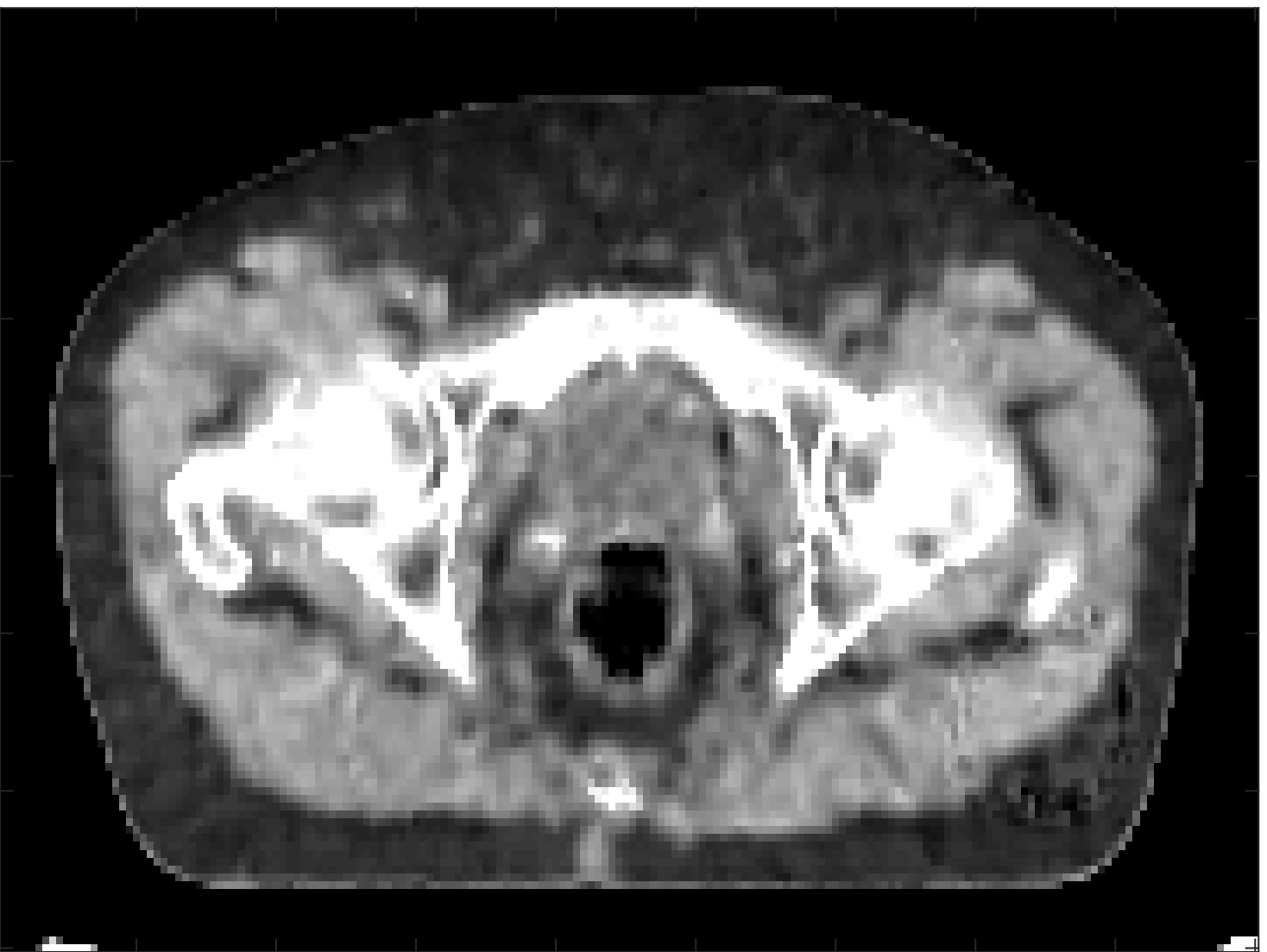}\\
			TV&
			{FBPConvNet}\\
			\includegraphics[width=.45\linewidth,height=.3\linewidth]{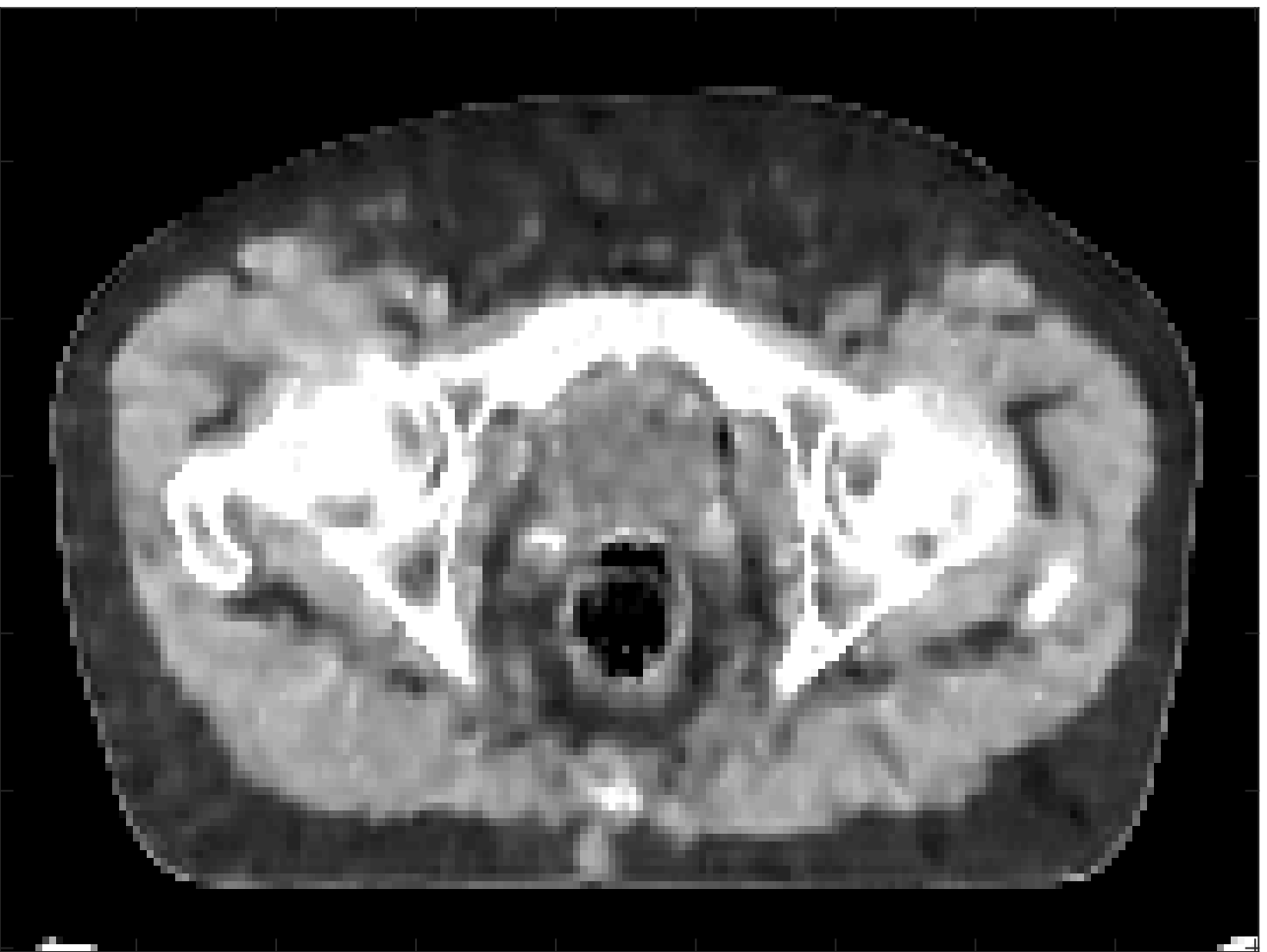}&
			\includegraphics[width=.45\linewidth,height=.3\linewidth]{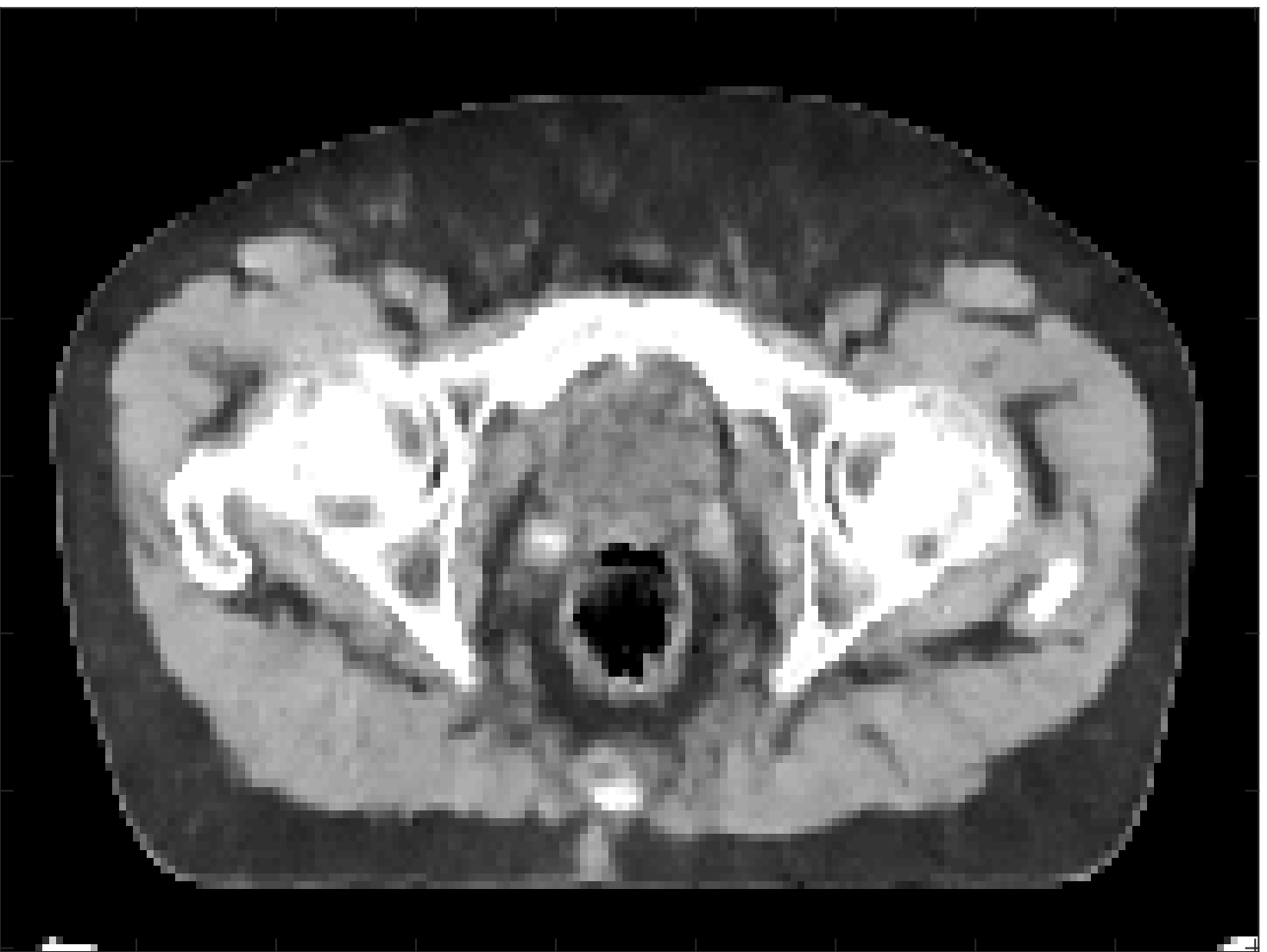}\\	
			PFBS-IR&
			PFBS-AIR
		\end{tabular}
		\caption {Reconstruction results at dose level $I_i=5\times10^3$.}
		\label{slice11rec_34_5000}
	\end{center}
\end{figure}

\begin{figure}[htb]
	\begin{center}
		\begin{tabular}{c@{\hspace{2pt}}}
			\includegraphics[width=.96\linewidth,height=.16\linewidth]{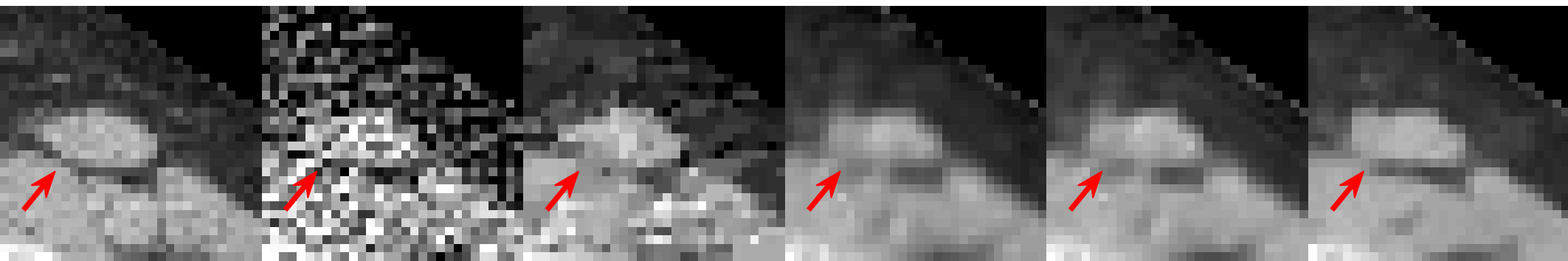}\\
			\includegraphics[width=.96\linewidth,height=.16\linewidth]{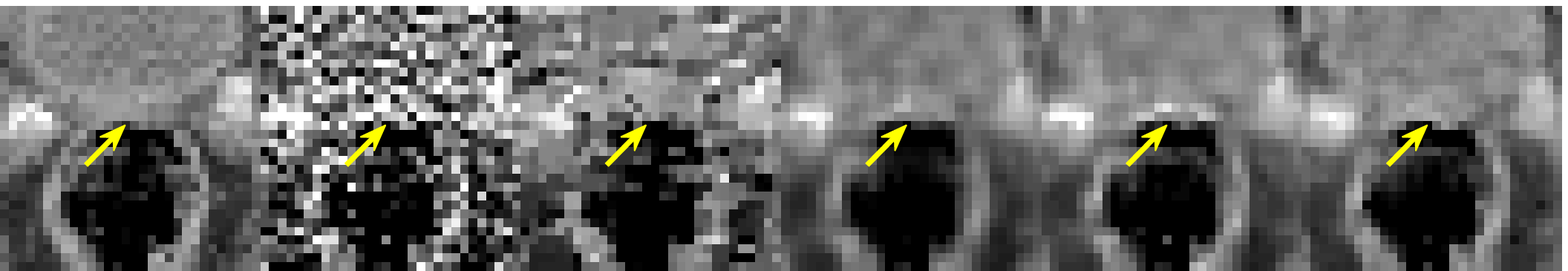}\\
			\includegraphics[width=.96\linewidth,height=.16\linewidth]{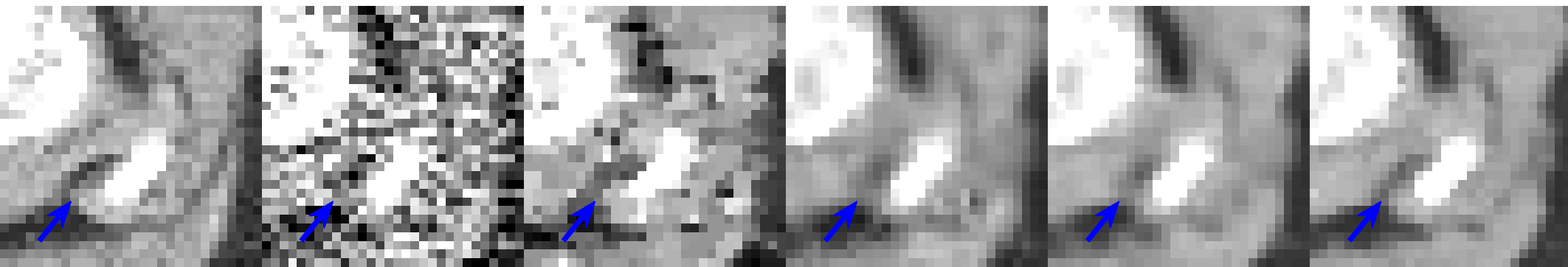}\\
		\footnotesize{~~~~NDCT~~~~~~FBP~~~~~~~~~~~TV~~~~~FBPConvNet~PFBS-IR~~~PFBS-AIR~~}
		\end{tabular}
		\caption{Zoom-in reconstruction results at dose level  $I_i=5\times10^3$.  Three rows from up to bottom correspond to the red, yellow and blue  boxes in Fig. \ref{slice11rec_34_5000} respectively, with differences highlighted in arrows.}
		\label{Zoomslice11rec_34_5000}
	\end{center}
\end{figure}

With further reduced dose,   Fig. \ref{slice12rec_38_10000} and Fig. \ref{slice11rec_34_5000} show images reconstructed
with  dose level of  $I_i=10^4$ and  $I_i=5\times10^3$, respectively.
And the corresponding
zoomed-in images  are displayed in Fig.   \ref{Zoomslice12rec_38_10000} and Fig. \ref{Zoomslice11rec_34_5000}.
These Figures suggest PFBS-AIR once again had the best reconstruction quality.



On the other hand, the quantitative results corresponding to Fig. \ref{slice2rec_36_50000} , Fig.  \ref{slice12rec_38_10000}, and Fig. \ref{slice11rec_34_5000} are listed in Table \ref{Tab:slice2rec_36}, Table \ref{Tab:slice12rec_38} and Table \ref{Tab:slice11rec_34} respectively, which also shows the best performance if PFBS-AIR in terms of PSNR, RMSE, and SSIM.

\begin{table}[htb]
	\centering
	\caption{Quantitative reconstruction results for the image slice in  Fig. \ref{slice2rec_36_50000}. }
	\scalebox{0.75}{
		\begin{tabular}{p{1.2cm}p{0.7cm}p{1.0cm}p{1.2cm}p{1.2cm}p{1.2cm}p{1.3cm}}
			\hline
			\hline   			
			{Dose level}                    &            & FBP          & TV          & FBPConvNet  &  PFBS-IR  & PFBS-AIR \\
			\hline   
			\multirow{3}{*}{$I_i=10^5$}     &PSNR        &   $42.4985$  &     $45.9763$    &$47.8269$    &$46.8933$         &    $\bold{51.5252}$          \\
			                                &RMSE        &   $~~0.0031$ &     $~~0.0021$   & $~~0.0019$  &$~~0.0021$        &    $~~\bold{0.0013}$          \\
		                                	&SSIM        &   $~~0.9936$ &     $~~0.9972$   & $~~0.9976$  &$~~0.9974$        &    $~~\bold{0.9986}$          \\
			\hline    
			\multirow{3}{*}{$I_i=5\times10^4$}&PSNR        &   $41.2469$  &     $44.4369$    & $46.8502$   & $45.0650$        &    $\bold{50.6984}$          \\
			                                &RMSE        &   $~~0.0036$ &     $~~0.0026$   & $~~0.0021$  &$~~0.0026$        &    $~~\bold{0.0014}$          \\
			                                &SSIM        &   $~~0.9904$ &     $~~0.9953 $  & $~~0.9966 $ &$~~0.9949$        &    $~~\bold{0.9984}$         \\
			\hline
			\multirow{3}{*}{$I_i=10^4$}     &PSNR        &   $36.3060$  &     $40.4774$    &$43.8588$    &  $43.2313$       &    $\bold{45.3091}$          \\
			                                &RMSE        &   $~~0.0066$ &     $~~0.0041$   &$~~0.0029$   &$~~0.0031$        &    $~~\bold{0.0022}$          \\
			                                &SSIM        &   $~~0.9638$ &     $~~0.9874$   &$~~0.9940$   &$~~0.9921$        &    $~~\bold{0.9963}$          \\
			\hline			
			\multirow{3}{*}{$I_i=5\times10^3$}&PSNR        &   $33.3903 $ &     $38.8149$    &$41.3073$    &$42.5498$         &    $\bold{45.1596}$          \\
			                                &RMSE        &   $~~0.0094$ &     $~~0.0051$   &$~~0.0036$   &$~~0.0034$        &    $~~\bold{0.0027}$          \\
			                                &SSIM        &   $~~0.9280$ &     $~~0.9816$   & $~~0.9912$  &$~~0.9913$        &    $~~\bold{0.9944}$          \\
			\hline                                                                                                                       
			\hline
		\end{tabular}
	}
	\label{Tab:slice2rec_36}
\end{table}

\begin{table}[htb]
	\centering
	\caption{Quantitative reconstruction results for the image slice in  Fig.  \ref{slice12rec_38_10000}.}
	\scalebox{0.75}{
		\begin{tabular}{p{1.2cm}p{0.7cm}p{1.0cm}p{1.2cm}p{1.2cm}p{1.2cm}p{1.3cm}}
			\hline
			\hline   
			
			{Dose level}                    &            & FBP         & TV            & FBPConvNet     &  PFBS-IR  & PFBS-AIR\\
			\hline   
			\multirow{3}{*}{$I_i=10^5$}     &PSNR        &   $39.0351$  &     $42.8197$         &$44.7526$    &$44.0438$        &    $\bold{48.1220}$          \\
			                                &RMSE        &   $~~0.0038$ &     $~~0.0025$        & $~~0.0020$  &$~~0.0022$       &    $~~\bold{0.0014}$          \\
			                                &SSIM        &   $~~0.9930$ &     $~~0.9832$        & $~~0.9966$  &$~~0.9970$       &    $~~\bold{0.9985}$          \\
			\hline    
			\multirow{3}{*}{$I_i=5\times10^4$}&PSNR        &   $38.1533$  &     $41.6874$         & $43.4913$   &$41.3381$        &    $\bold{46.6175}$          \\
			                                &RMSE        &   $~~0.0042$ &     $~~0.0028$        & $~~0.0023$  &$~~0.0029$       &    $~~\bold{0.0016}$          \\
			                                &SSIM        &   $~~0.9903$ &     $~~0.9887$        & $~~0.9960 $ &$~~0.9937$       &   $~~\bold{ 0.9980} $         \\
			\hline
			\multirow{3}{*}{$I_i=10^4$}     &PSNR        &   $34.6438$  &     $38.3544$         &$40.7496$    &$40.0446$        &    $\bold{43.0824}$          \\
		                                	&RMSE        &   $~~0.0064$ &     $~~0.0042$        &$~~0.0031$   &$~~0.0034$       &    $~~\bold{0.0024}$          \\
			                                &SSIM        &   $~~0.9689$ &     $~~0.9954$        &$~~0.9928$   &$~~0.9911$       &    $~~\bold{0.9955}$          \\
			\hline			
			\multirow{3}{*}{$I_i=5\times10^3$}&PSNR        &   $32.0027$  &     $36.0595$         &$38.8226$    &$39.2485$        &    $\bold{41.0346}$          \\
			                                &RMSE        &   $~~0.0087$ &     $~~0.0052$        &$~~0.0038$   &$~~0.0036$       &    $~~\bold{0.0029}$          \\
		                                	&SSIM        &   $~~0.9415$ &     $~~0.9968$        & $~~0.9899$  &$~~0.9903$       &    $~~\bold{0.9937}$          \\
			\hline                                                                                                                       
			\hline
		\end{tabular}
	}
	\label{Tab:slice12rec_38}
\end{table}
\begin{table}[htb]
	\centering
	\caption{Quantitative results for the image slice in Fig. \ref{slice11rec_34_5000}.}	
	\scalebox{0.75}{
		\begin{tabular}{p{1.2cm}p{0.7cm}p{1.0cm}p{1.2cm}p{1.2cm}p{1.2cm}p{1.3cm}}
			\hline    
			\hline   
			{Dose level}                    &            & FBP         & TV            & FBPConvNet     &  PFBS-IR  & PFBS-AIR\\
			\hline
			\multirow{3}{*}{$I_i=10^5$}     &PSNR        &   $39.1532$  &     $42.1400$        &$43.4265$    &$42.6736$        &    $\bold{46.3518}$          \\
		                                 	&RMSE        &   $~~0.0036$ &     $~~0.0025$       & $~~0.0023$  &$~~0.0025$       &    $~~\bold{0.0016}$          \\
		                                	&SSIM        &   $~~0.9925$ &     $~~0.9961$       & $~~0.9963$  &$~~0.9961$       &    $~~\bold{0.9978}$          \\
			\hline    
			\multirow{3}{*}{$I_i=5\times10^4$}&PSNR        &   $37.9938$  &     $40.7044$        & $42.3560$   &$40.6097$        &    $\bold{44.9094}$          \\
			                                &RMSE        &   $~~0.0041$ &     $~~0.0030$       & $~~0.0025$  &$~~0.0031$       &    $~~\bold{0.0019}$          \\
		                                	&SSIM        &   $~~0.9889$ &     $~~0.9938$       & $~~0.9949 $ &$~~0.9934$       &    $~~\bold{0.9970}$          \\
			\hline
			\multirow{3}{*}{$I_i=10^4$}     &PSNR        &   $33.4855$  &     $36.6776$        &$39.7747$    &$39.2662$        &    $\bold{41.4380}$          \\
			                                &RMSE        &   $~~0.0072$ &     $~~0.0048$       &$~~0.0035$   &$~~0.0037$       &    $~~\bold{0.0028}$          \\
			                                &SSIM        &   $~~0.9588$ &     $~~0.9839$       &$~~0.9913$   &$~~0.9896$       &    $~~\bold{0.9939}$          \\
			\hline
			\multirow{3}{*}{$I_i=5\times10^3$}&PSNR        &   $30.1080$  &     $35.0597$        &$37.4600$    &$37.7930$        &    $\bold{39.8136}$          \\
			                                &RMSE        &   $~0.0102$  &     $~~0.0058$       &$~~0.0042$   &$~~0.0041$       &    $~~\bold{0.0033}$          \\
			                                &SSIM        &   $~~0.9185$ &     $~~0.9763$       & $~~0.9879$  &$~~0.9880$       &    $~~\bold{0.9919}$          \\
			\hline                                                                                                                       
			\hline
		\end{tabular}
	}
	\label{Tab:slice11rec_34}
\end{table}

\section{Conclusion}
\label{concludesion}
We have developed a DL-regularized image reconstruction method for LDCT, using the optimization framework of PFBS, with (A)IR for preconditioned data-fidelity update, namely PFBS-(A)IR. The preliminary results suggest PFBS-AIR had superior reconstruction quality over FBP (an AR method), TV (an IR method), FBPConvNet (a DL-based image postprocessing method), and PFBS-IR (a DL-regularized image reconstruction method),  owing to the synergistic integration of AR, IR, and DL for LDCT.
\IEEEtriggeratref{50}
\bibliographystyle{ieeetr}
\bibliography{biblistS}
%
%
\end{document}